\documentclass[reprint,superscriptaddress,floatfix,amsmath,amssymb,pra,aps,showpacs,twocolumn]{revtex4-2}
\usepackage{amsmath,amsfonts,amssymb,amsthm,graphics,graphicx,epsfig,bbm}
\usepackage[colorlinks=true,citecolor=blue,linkcolor=blue,urlcolor=blue]{hyperref}
\usepackage[usenames]{color}
\usepackage{graphicx}
\usepackage[export]{adjustbox}
\usepackage{subfigure}
\usepackage{amsmath}
\usepackage{epsfig}
\usepackage{dcolumn}
\usepackage{bm}
\usepackage{color}
\usepackage{times}
\usepackage{epstopdf}
\usepackage{amssymb}
\usepackage{amstext}
\usepackage{latexsym}
\usepackage{hyperref}
\usepackage{amsfonts}
\usepackage{psfrag}
\usepackage[utf8]{inputenc}
\usepackage{graphicx}
\usepackage{amsmath}
\usepackage{amssymb}
\usepackage{dsfont}
\usepackage{tensor}
\usepackage{color}
\usepackage{ulem}
\usepackage{upgreek}
\usepackage{braket}
\usepackage{algpseudocode}
\usepackage{algorithm}
\usepackage{orcidlink}  

\UseRawInputEncoding

\newcommand{\supervec}[1]{\!\vec{\hskip 0.03em #1}\hskip 0.03em} 

\begin{document}

\title{Magnetic and phonon-induced effects on the non-Markovian dynamics of a single solid-state defect}

\author{Ariel Norambuena\orcidlink{0000-0001-9496-8765}}
\email{ariel.norambuena@umayor.cl}
\affiliation{Centro Multidisciplinario de F\'isica, Universidad Mayor, ´
Camino la Piramide 5750, Huechuraba, Santiago, Chile}

\author{Diego Tancara\orcidlink{0000-0002-5053-3521}}
\affiliation{Centro Multidisciplinario de F\'isica, Universidad Mayor, ´
Camino la Piramide 5750, Huechuraba, Santiago, Chile}
\affiliation{Departamento de F\'isica, Universidad de Santiago de Chile (USACH), Avenida V\'ictor Jara 3493, 9170124, Santiago, Chile}

\author{Vicente Chomal\'i-Castro\orcidlink{0009-0002-7218-1245}}
\affiliation{Department of Physics, University of Illinois Urbana-Champaign,\\
1110 West Green Street, Urbana, IL 61801, USA}

\author{Daniel Castillo\orcidlink{0000-0002-9556-4831}}
\affiliation{Centro Multidisciplinario de F\'isica, Universidad Mayor, ´
Camino la Piramide 5750, Huechuraba, Santiago, Chile}

\date{\today}

\begin{abstract}

The electron-phonon interaction is one of the most fundamental mechanisms in condensed matter physics. Phonons can induce memory effects in solid-state platforms when localized electronic states interact with lattice vibrations in non-unitary dynamical maps. In this work, we demonstrate how single-mode and structured phonon environments can give rise to non-Markovian dynamics of an individual negatively charged silicon-vacancy center in diamond. Using trace distance as a quantifier via numerical simulations and theoretical calculations, we identify the physical conditions for emerging and understanding non-Markovian behavior in diverse scenarios. Most importantly, we investigate the influence of magnetic fields (longitudinal and transverse), phonon couplings, Fock states, and temperature to understand how these factors influence memory effects in this solid-state device.
\end{abstract}

\maketitle

\section{Introduction}
Quantum non-Markovianity (NM) has emerged as an active research field, addressing both fundamental questions and novel applications in open quantum systems~\cite{Breuer2016}, biological structures~\cite{Nori2015}, and biochemical reactions~\cite{Torella2015}. In quantum technologies, non-Markovian dynamics can be harnessed to improve performance through control protocols, error correction, and decoupling pulses. This opens the door to advancements in quantum metrology~\cite{Plenio2012}, quantum communication channels~\cite{Maniscalco2014}, entanglement protocols~\cite{Mirkin2019}, and quantum control techniques~\cite{Koch2015}. Additionally, significant theoretical progress has been made in defining NM, particularly in analyzing its effects on generic open quantum systems~\cite{BLP2009,LPB2010,Luo2012,RHP2010,RHP2014}. Understanding NM is essential for identifying new platforms where information backflow can be tested, controlled, and applied in current quantum technologies. \par

In recent years, solid-state devices with structured phonon environments, such as nitrogen-vacancy (NV) and silicon-vacancy (SiV) centers in diamond, have been successfully implemented for applications like single-photon sources~\cite{Naydenov2014, Beratos2002, Wang2005, Sipahigil2014}, biological imaging~\cite{Fu2007, Fuchs2011, Lukin2008, Higbie2017}, spin-qubit quantum information processing~\cite{Ajoy2015}, and nanomechanical resonators for cooling schemes~\cite{Rabl2009, Kepesidis2013, Kepesidis2016}. These platforms also hold potential for exploring phonon-mediated non-Markovian dynamics~\cite{Haase2018, Ariel2020}, offering new perspectives to study memory effects in open quantum systems. In particular, the negatively charged silicon-vacancy (SiV$^-$) center in diamond stands out due to its strong spin-orbit interaction~\cite{Hepp2014}, which influences the coupling between phonons and the residual spin, enabling interesting phonon-orbital-spin dynamics that could be exploited for applications such as phonon-based quantum networks~\cite{Lemonde2018}, strain-induced control~\cite{Meesala2018}, quantum information processing~\cite{Hanks2020}, coherent interaction between defect~\cite{Day2022}, and preparation of squeezed states~\cite{Chen2021}. \par

In this work, we investigate phonons --- single-mode or structured phonon bath --- can induce non-Markovian behavior in the SiV$^{-}$ center. First, we focus on the SiV$^-$ center in diamond, modeling its non-Markovian dynamics induced by a single phonon mode. In this case, we analyze the role of the longitudinal magnetic field and phonon coupling to show how the SiV$^{-}$ center can be modeled using an effective Rabi model. Using this effective model, we derive approximate results to understand the role of Fock states and magnetic field strength. Then, we incorporate the transverse magnetic field to show how complex patterns can emerge by numerically computing the Breuer-Laine-Piilo non-Markovian measure. Finally, we focus on the dynamic induced by a structured phonon environment, where localized interactions are motivated by a diamond structure with circular holes. \par

\begin{figure}[ht!]
\centering
\hbox{\hspace{0.3 em} \includegraphics[width = 0.9 \linewidth]{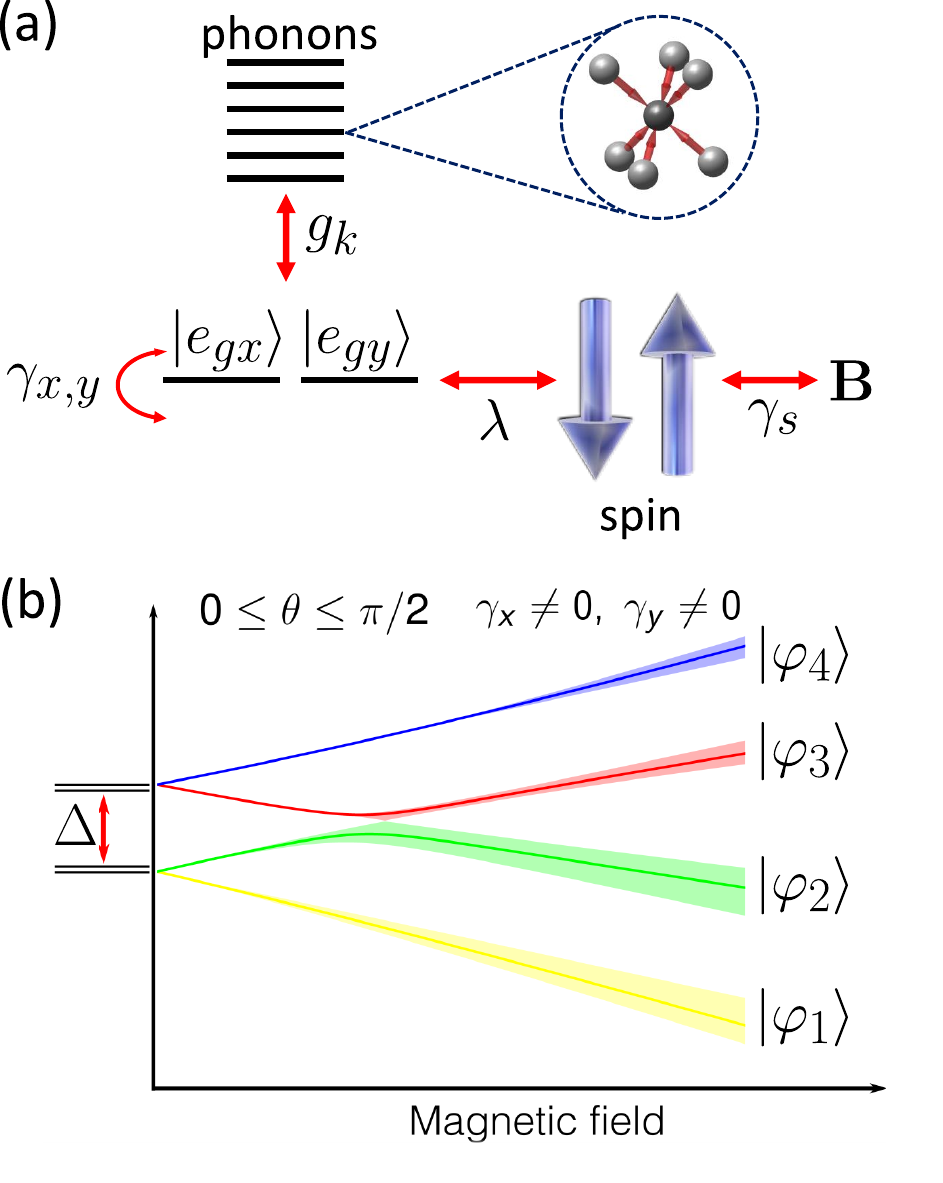}}
\caption{(a) Schematic representation of the main components of the SiV$^{-}$ center Hamiltonian and its interaction with phonons, showing electron-phonon couplings $g_k$, Jahn-Teller effect $\gamma_{x,y}$, spin-orbit interaction $\lambda$, and external magnetic field $\mathbf{B}$. (b) Energy levels of the SiV$^{-}$ as a function of the magnetic field strength $B = |\mathbf{B}|$. The filled colored areas represent all possible values of the eigenenergies $E_i$ for $0 \leq \theta \leq \pi/2$, $\Delta/2\pi=(\lambda^2+\gamma_x^2+\gamma_y^2)^{1/2} = 48$ GHz. }
\label{fig:Figure1}
\end{figure} 

\section{Negatively charged silicon-vacancy center in diamond}

The ground state of a SiV$^-$ center in diamond consists of two degenerate orbital states, $|e_{gx}\rangle$ and $|e_{gy}\rangle$, coupled to a residual electronic spin $S = 1/2$. The dynamics of this point defect under an external magnetic field $\mathbf{B}$ is governed by the following Hamiltonian~\cite{Kepesidis2016}
\begin{eqnarray}\label{Hamiltonian} 
H_{\rm SiV} = -\lambda L_z S_z + H_{\rm JT} + f \gamma_{\rm L} B_z L_z + \gamma_s \mathbf{B} \cdot \mathbf{S}, 
\end{eqnarray}
where $\lambda > 0$ is the spin-orbit coupling constant, $L_z$ is the $z$ component of the orbital angular momentum operator, $\mathbf{S} = (\hbar/2) \boldsymbol{\sigma}$ represents the spin operator, with $\sigma_{\alpha}$ denoting the Pauli matrices for $S = 1/2$. The Hamiltonian $H_{\rm JT} = (1/2)(\gamma_x \sigma_z - \gamma_y \sigma_x)$ accounts for the Jahn-Teller (JT) effect, where $\gamma_{x,y}$ are the JT coupling constants~\cite{Bersuker2006}. The orbital gyromagnetic ratio is given by $\gamma_L = \gamma_s/2$, with $f \approx 0.1$ being the Ham reduction factor, and $\gamma_s = 2.8$ MHz/G represents the electronic spin gyromagnetic ratio. The orbital operators are defined as $\sigma_x = |e_{gx}\rangle \langle e_{gy}| + |e_{gy}\rangle \langle e_{gx}|$, $\sigma_y = -i|e_{gx}\rangle \langle e_{gy}| + i|e_{gy}\rangle \langle e_{gx}|$, and $\sigma_z = |e_{gx}\rangle \langle e_{gx}| - |e_{gy}\rangle \langle e_{gy}|$. Fig.~\ref{fig:Figure1}(a) provides a schematic representation of the internal dynamics of the SiV$^{-}$ system, illustrating phonon-induced effects on the orbital states. \par

Orbital states with symmetries different from the $E$ symmetry are energetically far from the degenerate states $|e_x\rangle$ and $|e_y\rangle$, implying that $L_x = L_y = 0$ for the ground state manifold of the SiV$^{-}$ center~\cite{Hepp2014}. Furthermore, due to symmetry considerations we have that $L_z = \hbar \sigma_y$, with $L_z |e_{\pm}\rangle = \pm \hbar |e_{\pm}\rangle$, where $|e_{\pm}\rangle = (|e_{gx}\rangle \pm i |e_{gy}\rangle)/\sqrt{2}$ are the normalized orbital eigenstates of the spin-orbit interaction. Experimental measurements indicate that $\lambda/2\pi = (48-50)$ GHz, $\gamma_x/2\pi = (1-2)$ GHz, and $\gamma_y/2\pi = (2-3)$ GHz~\cite{Hepp2014}, confirming that spin-orbit coupling dominates at zero magnetic field. This point defect has a weak JT interaction since $\Upsilon = (\gamma_x^2 + \gamma_y^2)^{1/2} \ll \lambda$. \par

For a longitudinal magnetic field $\mathbf{B} = (0,0,B_z)$, the eigenstates of the SiV$^{-}$ center can be approximately described as $\ket{1} \approx |e-,\downarrow\rangle$, $\ket{2} \approx |e_+,\uparrow\rangle$, $\ket{3} \approx |e_+,\downarrow\rangle$, and $\ket{4} \approx |e_-,\uparrow\rangle$~\cite{Kepesidis2016, Lemonde2018}. The effect of a generalized magnetic field $\mathbf{B} = B(\cos(\phi)\sin\theta,\sin(\phi)\sin\theta,\cos(\theta))$ on the energy levels $E_i$ of the SiV$^{-}$ is illustrated in Fig.~\ref{fig:Figure1}(b). When lattice phonons are considered, the SiV$^{-}$ center exhibits the spin-preserving transitions $|1\rangle \leftrightarrow |3\rangle$ (spin $\downarrow$) and $|2\rangle \leftrightarrow |4\rangle$ (spin $\uparrow$). However, when transverse magnetic field components ($B_x$ or $B_y$) are taken into account, new transitions between states $|1\rangle \leftrightarrow |4\rangle$ and $|2\rangle \leftrightarrow |3\rangle$ appears. Therefore, the role of the magnetic field is crucial to understanding the non-Markovian response in this solid-state state. Additionally, temperature plays a key aspect in phonon baths at thermal equilibrium. All these ingredients make the SiV$^{-}$ center a challenging four-level to study in the context of NM.\par 

In the next section, we will introduce a physically motivated study of the non-Markovian dynamics of the SiV$^{-}$ center. In particular, we first focus on the physics of a single phonon mode to understand the role of the magnetic field, phonon coupling, and Fock states via numerical simulations and some approximate analytical solutions.

\section{Phonon-induced non-Markovianity induced by a single phonon mode} 

To understand the fundamental phonon-induced non-Markovian physics, let us consider the dynamics of a SiV$^{-}$ center coupled to a single phonon mode. This scenario can be described using the Hamiltonian ($\hbar = 1$)
\begin{equation} \label{SingleModeHamiltonian}
    H = H_{\rm SiV} + H_{\rm ph} + H_{\rm SiV-ph},
\end{equation}
where $H_{\rm SiV}$ is the SiV$^{-}$ Hamiltonian given in Eq.~\eqref{Hamiltonian}, $H_{\rm ph} = \omega_{\rm ph} c^{\dagger}c$ is the phonon Hamiltonian of the single mode, $\omega_{\rm ph}$ is the phonon frequency, $H_{\rm SiV-ph} = g_1 (c^{\dagger}+c)(L_-+L_+)-ig_2(c^{\dagger}+c)(L_- - L_+)$ is the interacting Hamiltonian~\cite{Kepesidis2016}, $g_{1,2}$ are the phonon couplings, and $L_+ = \ket{3}\!\bra{1}+\ket{2}\!\bra{4} = L_-^{\dagger}$ is the orbital raising operator ($L_+ \ket{e_-} = \ket{e_+}$). The interaction Hamiltonian used here is derived from group theoretical calculations without applying the rotation wave approximation (RWA); see Appendix B of Ref.~\cite{Kepesidis2016} for more details. The latter is relevant since, as explained in Ref.~\cite{Mäkelä2013}, the RWA can reduce the observed non-Markovianity in a quantum system.  \par

For compression phonon modes propagating along some axis, it has been demonstrated that the phonon coupling constant is approximately given by $g_{\rm ph} \approx d/v_s(\hbar \omega_{\rm ph}/(2 \rho l w t))^{1/2}$~\cite{Nori2019}, where $d/2\pi \sim 10^{15}$ Hz$/$strain is the strain sensitivity, $v_s = 1.2 \times 10^4$ m/s is the speed of sound in diamond, and $\rho = 3500$ kg/m$^3$ is the diamond density. Suppose we have a resonant phonon mode $\omega_{\rm ph} \approx \Delta$, where $\Delta = (\lambda^2+\Upsilon^2)^{1/2} \approx 2\pi \times 50$ GHz is the typical energy gap of the SiV$^{-}$ center at zero magnetic field. In a cantilever with dimensions $(l, w, t) = (25, 0.1, 0.1) \mu$m~\cite{Kepesidis2016} one obtains a single mode coupling $g_{\rm ph}^{\rm cant}
/2\pi \approx 16$ MHz. Thus, if we use the values reported for the phononic crystal with holes~\cite{Nori2019}, where each SiV$^{-}$ is placed between holes, we obtain $g_{\rm ph}^{\rm phon,crys}/2\pi \approx 2.6$ GHz. Based on these estimations, $g_{\rm ph}^{\rm cant}/\Delta \approx 3.2 \times 10^{-4}$ and $g_{\rm ph}^{\rm phon,crys}/\Delta \approx 0.05$, we can assume for our simulations that the phonon coupling constant will be approximately $g_{\rm ph} \sim 10^{-4} -10^{-1}\Delta$. Strategies to engineer a larger single phonon coupling are: (i) enhancing the strain sensitivity~\cite{Meesala2018} or (ii) designing a phononic crystal with dimensions $(l, w, t)$ in the nano regime~\cite{Nori2019} to create strong defect-phonon interactions with localized vibrations. \par

Under realistic conditions, a single phonon mode will decay with a damping rate $\gamma_{\rm ph} = \omega_{\rm ph}/Q$, where $Q$ is the mechanical quality factor, whose expected values are $Q \approx 10^5-10^6$~\cite{Kepesidis2016}. For the case $\omega_{\rm ph} \approx \Delta$, we obtain $g_{\rm ph} \gg \gamma_{\rm ph}$. The phonon dissipation effect is always present since any single phonon mode travels to some edge, causing dissipation of the vibrational mode. In addition, we must also include phonon-induced dissipation on the SiV$^{-}$ center to perform realistic simulations. Then, taking into account all dissipation channels, we can phenomenologically model the dynamics using the following master equation in the Lindblad form ($\hbar = 1$)
\begin{equation} \label{MasterEquationSiVSingleMode}
    \dot{\rho} = -i[H,\rho] + \mathcal{L}_{\rm ph}(\rho) + \mathcal{L}_{\rm SiV}(\rho),
\end{equation}
where $\mathcal{L}_{\rm ph}(\rho) = \gamma_{\rm ph}(N(\omega_{\rm ph})+1)\mathcal{D}_{c}(\rho) + \gamma_{\rm ph} N(\omega_{\rm ph})\mathcal{D}_{c^{\dagger}}(\rho)$ takes into account the phonon dissipation with $N(\omega_{\rm ph}) = [{\mbox{exp}(\hbar \omega_{\rm ph}/k_{\rm B}T)-1}]^{-1}$ representing the mean number of phonons according to the Bose-Einstein distribution. Here, $\mathcal{D}_{O}(\rho) = O \rho O^{\dagger}-(1/2)\{O^{\dagger}O, \rho\}$ is the usual operator of the Lindblad master equation. The second term on Eq.~\eqref{MasterEquationSiVSingleMode} is the SiV$^{-}$ center relaxation dynamics, where $\mathcal{L}_{\rm SiV}(\rho) = \Gamma_{\rm SiV} (N(\Delta)+1)\mathcal{D}_{J_-}(\rho) + \Gamma_{\rm SiV}  N(\Delta)\mathcal{D}_{J_+}(\rho)$ and $N(\Delta)$ is the mean number of phonons at the SiV$^{-}$ energy gap. In our numerical simulations, we use the reported value $\Gamma_{\rm SiV}/2\pi \approx 1.78$ MHz~\cite{Kepesidis2016}, which is a good approximation at low temperatures (few kelvins). As known in the color center community, the SiV$^{-}$ relaxation rates have a temperature dependence for temperatures ranging from a few millikelvins to room temperature~\cite{Sipahigil2014}. Here, we use the values at very low temperatures reported for the SiV center. Moreover, for a large value of the phonon coupling $g_{\rm ph} \sim 10^{-1} \Delta $, we can satisfy the condition $g_{\rm ph} \gg \Gamma_{\rm SiV}$. In summary, we have two important physical conditions in this model: $g_{\rm ph} \ll \Delta, \omega_{\rm ph}$ and $g_{\rm ph} \gg \Gamma_{\rm SiV} > \Gamma_{\rm ph}$. \par

We shall illustrate how phonon-induced non-Markovianity of the SiV$^{-}$ center can be modeled using an effective Rabi model when the magnetic field aligns with the symmetry axis of this point defect.

\subsection{Effective Rabi model for longitudinal magnetic field}

As a first example, we consider a longitudinal magnetic field $\mathbf{B} = (0,0, B_z)$ to demonstrate that the dynamics SiV$^{-}$ coupled to single mode can be modeled by the quantum Rabi model depending on the SiV$^{-}$ initial state. In the regime $\Upsilon \ll \lambda$, the eigenenergies of the SiV$^{-}$ center are given by
\begin{eqnarray}
E_1 &=& {1 \over 2}\left(-\gamma_s B_z - \Delta_- \right), \quad \ket{1} \approx |e_-,\downarrow\rangle \label{E1}\\
E_2 &=& {1 \over 2}\left(+\gamma_s B_z - \Delta_+ \right), \quad \ket{2} \approx |e_+,\uparrow\rangle \label{E2}\\
E_3 &=& {1 \over 2}\left(-\gamma_s B_z + \Delta_- \right), \quad \ket{3} \approx |e_+,\downarrow\rangle \label{E3}\\
E_4 &=& {1 \over 2}\left(+\gamma_s B_z + \Delta_+ \right), \quad \ket{4} \approx |e_-,\uparrow \rangle, \label{E4}
\end{eqnarray}
where $\Delta_{\pm} = [4(f \gamma_L B_z)^2 \pm 4 f \gamma_L B_z \lambda + \lambda^2 + \Upsilon^2]^{1/2}$. For a magnetic field satisfying $2 f \gamma_{\rm L} B_z \ll \lambda$ or equivalently $B_z \lesssim 10^2$ T (experimentally), we obtain $\Delta_{\pm} \approx \Delta = (\lambda^2 + \Upsilon^2)^{1/2} \approx 2\pi \times 50$ GHz~\cite{Hepp2014}, which is a good approximation for longitudinal magnetic fields $B_z \approx (0-10^2)$ T. In such a case, a single phonon mode with a frequency $\omega_{\rm ph} \approx \Delta$ can induce a resonant response on the SiV$^{-}$ center because $\Delta \approx E_{4}-E_{2} \approx E_{3}-E_1$. These energy transitions are relevant since the orbital raising operator in the SiV$^{-}$-phonon interaction Hamiltonian~\eqref{SingleModeHamiltonian} connects such states. \par

If the initial state of the SiV$^{-}$ belongs to one of the Hilbert sub-spaces $\mathcal{H}_1 : = \{\ket{1}, \ket{3}\}$ or $\mathcal{H}_2 := \{\ket{2}, \ket{4}\}$, the dynamics can be described by a simple two-level system coupled to single phonon mode. In any of these two Hilbert sub-spaces, we can define $\ket{g} = \{\ket{1} 
 \; \mbox{or} \; \ket{2}\}$ (ground state with energy $E_{g}$) and $\ket{e} = \{\ket{3} 
 \; \mbox{or} \; \ket{4}\}$ (excited state with energy $E_e$). Therefore, for each Hilbert space $\mathcal{H}_i$ ($i=1,2$), we found the following effective quantum Rabi Hamiltonian
\begin{equation} \label{EffH}
    H_{\rm eff} = {1 \over 2}\omega_{\rm s} S_z^{\rm eff} + \omega_{\rm ph} c^{\dagger}c +  (c+c^{\dagger})(g^{\ast} S_+^{\rm eff}+g S_-^{\rm eff}),
\end{equation}
where $\omega_{\rm s} = (E_e-E_g)/\hbar$, $g = g_1+ig_2$, $S_z^{\rm eff}  = \ket{e}\!\bra{e}-\ket{g}\!\bra{g}$, and $S_+^{\rm eff} = \ket{e}\!\bra{g} =(S_-^{\rm eff})^{\dagger}$. \par 

Now, we define the reduced density matrix of the SiV$^{-}$ center as $\rho_{\rm s}(t) = \mbox{Tr}_{\rm ph}(\rho)$, where we trace over the single phonon degree of freedom. We remark that for a longitudinal magnetic field, any signature of NM can be detected using the trace distance $D(t) = (1/2)\|\rho_s(t)-\rho_{\rm SS}\|$, where $\rho_{\rm SS}$ is the steady state for the SiV$^{-}$ center. Here, $\|X\| = \mbox{Tr}(\sqrt{X^{\dagger}X})$ denote the trace norm ($l_1$-norm). In the steady state, one obtains $\lim_{t \rightarrow \infty}D(t) = 0$, where possible oscillations in $D(t)$ can only be originated due to phonon interactions since the effective SiV$^{-}$ center Hamiltonian has no tunneling between states. Therefore, we are quantifying the indistinguishability between the instantaneous and steady-state of the SiV$^{-}$ center. As demonstrated in Ref.~\cite{Albarran2023_entropy} for a cavity-multi-qubit device with a multi-qubit Hamiltonian $H_{\rm qubit} = \sum_{i}^{}\omega_i S_z^i$, the trace distance $D(t)$ is a monotonic function in time if the system is Markovian. Thus, any signature of NM can be quantified in terms of violating the monotonicity of $D(t)$. This manner of quantifying NM is dynamical and does not depend on the initial state optimization used in the Breuer-Laine-Piilo (BLP) measure~\cite {BLP2009}. Therefore, for longitudinal magnetic fields, we can use the following dynamical non-Markovianity measure~\cite{Albarran2023_entropy}
\begin{equation} \label{DNM}
    \mathcal{N}_D = \int_{\dot{D}>0} \dot{D} \, dt, \quad D(t) = {1 \over 2}\|\rho_s(t)-\rho_{\rm SS}\|.
\end{equation}
\begin{figure}[h!]
\centering
\includegraphics[width = 0.9 \linewidth]{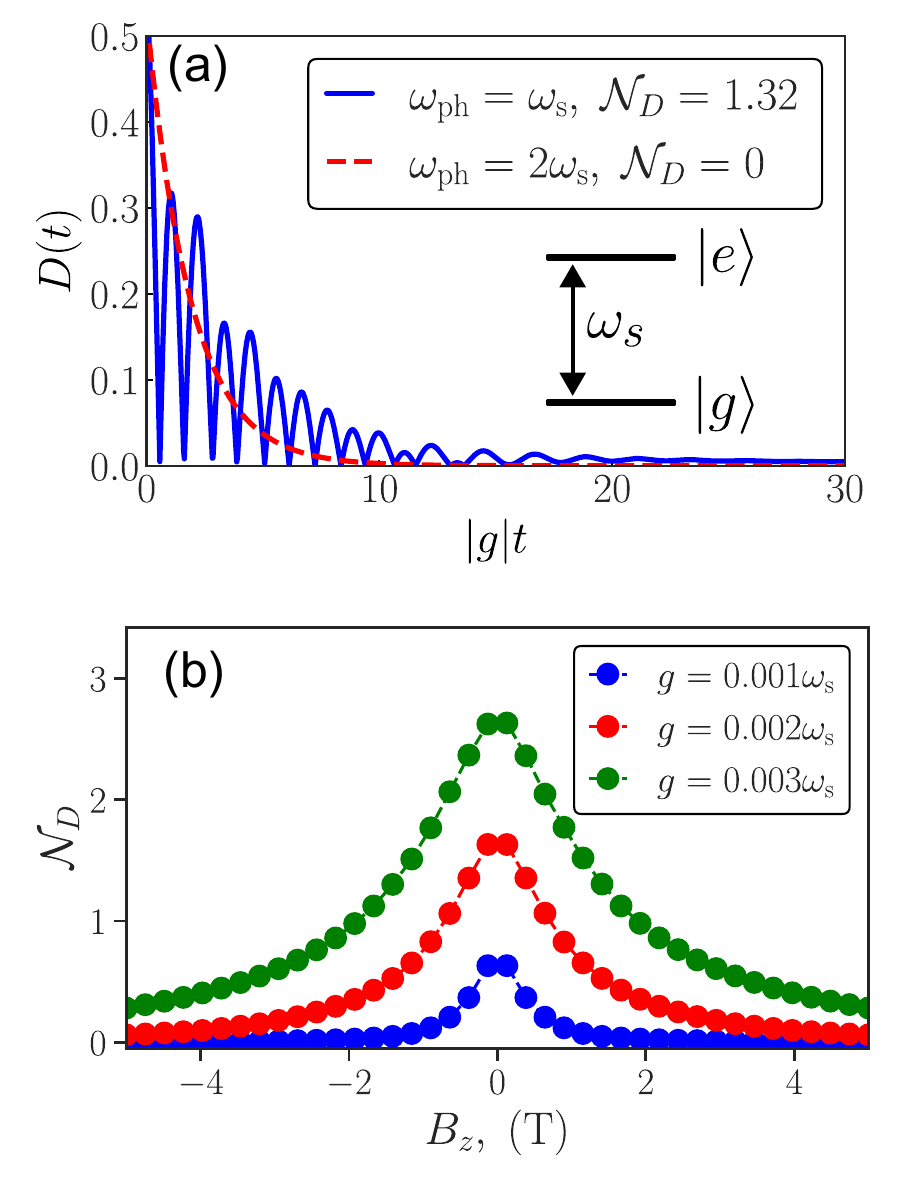}
\caption{(a) Dynamical behavior of the trace distance $D(t)$ for $\omega_{\rm ph} =  \omega_{\rm s}$ (solid blue) and $\omega_{\rm ph} = 2\omega_{\rm s}$ (red dashed), where $\omega_{\rm s} = (\lambda^2+\Upsilon^2)^{1/2}$ and $g= 10^{-3}\omega_{\rm s}$. (b) Dynamical degree of non-Markovianity as a function of the longitudinal magnetic field $B_z$ for different coupling values $g$. For the simulation we considered $\rho_{\rm SiV}(0) = \ket{1} = |e_-,\downarrow\rangle$ (ground state of the SiV$^{-}$), $\rho_{\rm ph}(0) = \ket{n}$ (Fock state with $n=0$), $\lambda/2\pi = 45$ GHz, $\gamma_{x,y} = 1$ GHz, $Q = 10^5$, $\gamma_{\rm ph} = \omega_{\rm ph}/Q$, $N(\omega_{\rm ph})=0$, $\Gamma_{\rm SiV} = 2 \pi \times 1.78 \times 10^6$ Hz, and $N(\Delta) = 10$.}
\label{fig:Figure2}
\end{figure}

Note that the integral in Eq.~\eqref{DNM} is calculated until the system reaches a steady state. In Fig.~\ref{fig:Figure2}(a), we illustrate the resonant case ($\omega_{\rm ph} = \omega_{\rm s}$) and off-resonant ($\omega_{\rm ph} = 2\omega_{\rm s}$) response of the trace distance $D(t)$. In addition, the dynamical non-Markovianity measure~\eqref{DNM} is calculated in both cases to illustrate the effect of the phonon frequency. The initial condition is fixed as $\ket{i} = \ket{1}$ and $|\Psi_{\rm ph}(0)\rangle =|n\rangle$ (Fock state with $n=1$). We observe that the trace distance $D(t)$ exhibits oscillations due to the coherent coupling with the resonant phonon mode, which is (for longitudinal fields) a signature of phonon-induced NM on the SiV${-}$ center. On the contrary, in the off-resonant case, we note that oscillations disappear, which is the usual Markovian response. \par
 
In addition, we can analyze the non-Markovian response of the system for different values of the longitudinal magnetic fields $B_z$. In Fig.~\ref{fig:Figure2}(b), we show the dynamical degree of NM as a function of $B_z$ and for different values of the phonon coupling $|g|$. According to our simulations, the maximum value of $\mathcal{N}_D$ scales linearly with the phonon coupling, and thus, $\mbox{max}[\mathcal{N}_D] \propto g$. Another interesting observation is the Lorentzian shape of $\mathcal{N}_D(B_z)$, indicating that the system can exhibit this symmetrical non-Markovian behavior even for a range of values magnetic fields $B_z$ around the maximum peak $B_z = 0$ (when $\omega_{\rm ph} = \omega_{\rm s}$). \par

The effective Hamiltonian~\eqref{EffH} is similar to the quantum Rabi model discussed in Refs.~\cite{Xie2017,Yu2012}, where the exact solution is analyzed in detail. Here, we can apply a similar procedure by considering the role of the phase $\theta = \mbox{atan}(g_2/g_1)$. In Ref.~\cite{Kepesidis2016}, it is assumed that $g_1 \approx g_2$ is of the order of PHz~\cite{Mamin2012}, which is equivalent to have $\theta = \pi/4$. By introducing the state $|\Psi\rangle = [\psi_1, \psi_2]^T$ in the basis $|e\rangle = [1, 0]^{T}$ and $|g\rangle = [0, 1]^{T}$, and after applying the eigenvalue equation $H_{\rm eff}|\Psi\rangle = E |\Psi\rangle$, we get
\begin{eqnarray}
    \omega_{\rm ph} c^{\dagger}c \psi_1 + |g|e^{-i\theta}(c^{\dagger}+c) \psi_2 + {\omega_{\rm s} \over 2} \psi_1 &=&E \psi_1, \\
    \omega_{\rm ph} c^{\dagger}c \psi_2 + |g|e^{i\theta}(c^{\dagger}+c) \psi_1 - {\omega_{\rm s} \over 2} \phi_2 &=&E \psi_2.
\end{eqnarray}
For the case $g_2=0$ (or $\theta = 0, 2\pi$), we obtain the same system of equations for the quantum Rabi model reported in Ref~\cite{Xie2017}. By defining $\phi_1 = e^{i\theta/2}\psi_1+e^{-i\theta/2}\psi_2$ and $\phi_2 = e^{i\theta/2}\psi_1-e^{-i\theta/2}\psi_2$, we obtain the eigenvalue equation
\begin{eqnarray}
   &&  H_R  \left(\begin{array}{c}
       \phi_1  \\
      \phi_2   
    \end{array} \right) = E\left(\begin{array}{c}
       \phi_1  \\
      \phi_2   
    \end{array} \right), \\
    H_R &= &\left(\begin{array}{cc}
      c^{\dagger}c + |g|(c^{\dagger}+c)  & \displaystyle{{\omega_{\rm s} \over 2}}\\
      \displaystyle{{\omega_{\rm s} \over 2}}  &  c^{\dagger}c - |g|(c^{\dagger}+c)
    \end{array} \right).
\end{eqnarray}

The above system can be solved using the Bogoliubov transformation or the Bargmann-Fock space. \par

In the regime $|g| \ll \omega_{\rm ph}, \omega_{\rm s}$ and $|g| \gg \Gamma_{\rm SiV}, \gamma_{\rm ph}$, we can apply the rotating wave approximation on the effective Rabi model~\eqref{EffH}, leading to the Jaynes-Cummings Hamiltonian $H_{\rm JC} = {1 \over 2}\omega_{\rm s} S_z^{\rm eff} + \omega_{\rm ph} c^{\dagger}c+ |g|(e^{i\theta} c S_+^{\rm eff}+e^{-i\theta}c^{\dagger}S_-^{\rm eff})$. This simple JC model is useful to explain why the resonant condition $\omega_{\rm ph} = \omega_{\rm s}$ leads to a maximal signature of non-Markovianity, as is also explained in Ref.~\cite{Albarran2023_entropy}. In simple words, the detuning $\delta = \omega_{\rm ph} -\omega_{\rm s}$ decides whether or not there is a non-Markovian response, and the phonon coupling impact on how fast or slow the oscillations of $D(t)$ are. \par

In the following subsection, we will analyze the role of the non-Markovian dynamics generated from different initial Fock states and derive some analytical results for the dynamic degree of NM.

\subsubsection{Non-Markovianity generated by Fock states}
Let us suppose that the initial state is given by $|\Psi(0)\rangle = |g,n\rangle$, where $\ket{g}$ is the SiV$^{-}$ ground state and $\ket{n}$ is the phonon Fock state. As discussed previously, the maximum achievable non-Markovianity is reached for the resonant case $\omega_{\rm ph} = \omega_{\rm s} =\Delta$. In the weak-coupling regime $|g| \ll \omega_s$ and $|g| \gg \Gamma_{\rm SiV}, \gamma_{\rm ph}$ (which is our case), we can apply the rotating-wave approximation and the exact dynamics can be analyzed in the manifold $\{|g,n\rangle,|e,n-1\rangle\}$. In the basis $|\phi_1\rangle = |g,n\rangle = [1,0]^T$ and $|\phi_2\rangle = |e,n-1\rangle = [0,1]^T$, the dissipative dynamics predicted by Eq.~\eqref{MasterEquationSiVSingleMode}, leads to 
\begin{eqnarray} 
    \dot{\rho}_{11} &=& - \Gamma \rho_{11} - \gamma_1 - i\Omega_n \rho_{21} + i \Omega_n^{\ast}\rho_{12}, \label{rho11}\\
    \dot{\rho}_{12} &=& -{\Gamma \over 2} \rho_{12} + 2i\Omega_n \rho_{11} - i\Omega_n, \label{rho12}
\end{eqnarray}
where $\Gamma = \gamma_{\rm ph}(2N(\omega_{\rm ph})+1) + \Gamma_{\rm SiV}(2N(\Delta)+1)$, $\gamma_1 = \gamma_{\rm ph}N(\omega_{\rm ph} + \Gamma_{\rm SiV}(N(\Delta)+1)$, and $\Omega_n = g\sqrt{n+1}$. Here, $\rho_{ij} =  \langle \phi_i|\rho| \phi_j\rangle$ are the matrix element of the density matrix in the manifold $|\phi_{i}\rangle \in \{|g,n\rangle,|e,n-1\rangle\}$. Since we are trying to understand the non-Markovian response in terms of the initial Fock state, we will focus on estimating the trace distance given in Eq.~\eqref{DNM}. In particular, the relation $D(t) = |\rho_{11}(t)-\rho_{11}^{\rm ss}|$ is theoretically derived, where $\rho_{11}^{\rm ss}$ is the steady state of $\rho_{11}(t)$. By setting $\dot{\rho}_{11} = \dot{\rho}_{12} = 0$ (steady state), we found that
\begin{equation}
    \rho_{11}^{\rm ss} = {4|\Omega_n|^2 - \gamma_1 \Gamma \over 8|\Omega_n|^2+\Gamma^2}.
\end{equation}

We note that in the regime $|\Omega_n| \gg \Gamma,\gamma_1$ the steady state converge to $\rho_{11}^{\rm ss} = 1/2$, as expected. Thus, using Eqs.~\eqref{rho11} and \eqref{rho12}, and after finding the exact solution for $\rho_{11}(t)$, we found the following analytical solution for the trace distance
\begin{equation}
    D(t) = |f(t)|e^{-\Gamma_0 t}, \quad f(t) = C_1 \cos(\mu t) + C_2 \sin(\mu t), 
\end{equation}
with $C_1 = \rho_{11}(0)-\rho_{11}^{\rm ss}$, $C_2 = (\Gamma_0 C_1-\gamma_1 - \Gamma \rho_{12}(0))/\mu$, $\Gamma_0 = 3\Gamma/4$, $\mu = [(2|\Omega_n|)^2-(\Gamma/4)^2]^{1/2}$. Because of the condition $|g| \gg \Gamma_{\rm SiV}, \gamma_{\rm ph}$ we can guarantee that $\mu$ is a real number since $4|\Omega_n|^2>(\Gamma/4)^2$. We obtain 
\begin{eqnarray} \label{AnalyticDyNM}
    \mathcal{N}_D &=& \int_{\dot{D}>0} \dot{D} \, dt \approx  {1 \over 4}e^{-\alpha_1 \Gamma_0 / \mu} \mbox{sech}\left( {\Gamma_0 \pi \over \mu} \right).
\end{eqnarray}

In Appendix~\ref{appendix:c}, we show more details about the above derivation. For instance, if we apply this simple model to the system parameters used in Fig.~\ref{fig:Figure2}(a), we get $\mathcal{N}_D \approx 0.98$, which is reasonable and in the order of magnitude of the real numerical value ($1.32$). In the regime $|g|\gg \Gamma_0$, $\mu = [(2|\Omega_n|)^2-(\Gamma/4)^2]^{1/2} \approx \sqrt{2}\sqrt{n+1}|g|$ and $\mbox{sech}(\Gamma_0 \pi /\mu) \approx \mu /(\Gamma_0 \pi)$, which tell us that (approximately) $\mathcal{N}_D \propto \sqrt{n+1}|g|/[2N(\Delta)+1])$. Note that $\Gamma_0 = 3\Gamma/4 \approx \Gamma_{\rm SiV}[2N(\Delta)+1]$ for high quality factors $Q \sim 10^5-10^6$ ($\gamma_{\rm ph} \ll \Gamma_0$). The latter scaling implies that the dynamical degree of NM increases with the phonon coupling, the value of $n$ (initial Fock state $|n\rangle$), but decreases with the value of $N(\Delta)$. Physically, $N(\Delta) = [\mbox{exp}(\Delta/k_B T)-1]^{-1}$ decreases by increasing temperature (dominant effect) or by increasing the energy gap $\Delta$, which can be done using the external magnetic field. \par

Therefore, to enhance the non-Markovian response with an initial Fock state, we must increase the phonon coupling $|g|$ with a larger Fock number $n$ or decrease temperature, maintaining fix the detuning $\delta = \omega_{\rm ph} -\omega_{\rm s}=0$. However, a mean-field approximation can be applied to understand the physics in the regime $n = \langle c^{\dagger} c \rangle \gg 1$.

\begin{figure}[h!]
\centering
\includegraphics[width = 1 \linewidth]{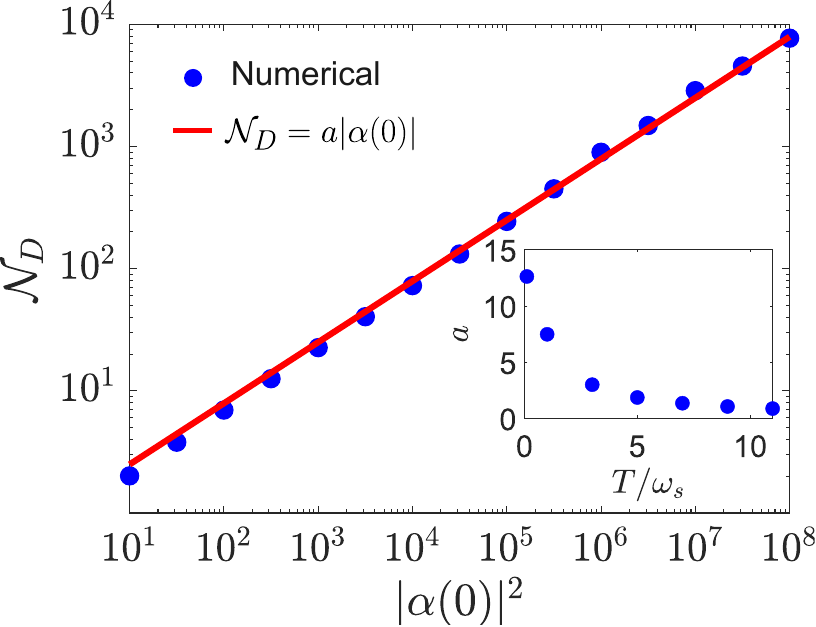}
\caption{Dynamical non-Markovianity as a function of the initial mean number of phonons under the mean field approximation. The fit line corresponds to the model $\mathcal{N}_D = a |\alpha(0)|$, where $a = 0.7898$ gives a statistical confidence bound ($0.7683, 0.8112$) at the 95\% confidence level. In the inset figure, we can observe the temperature dependence of the scaling factor $a$.}
\label{fig:Figure3}
\end{figure} 

\subsubsection{Non-Markovianity in the regime $\langle c^{\dagger} c\rangle  \gg 1$}
Let us consider the regime $n = \langle c^{\dagger} c\rangle \gg 1$, where we have a large mean number of phonons. In this regime, we can treat the mechanical mode as a classical field with amplitude $\alpha(t) = \langle c(t) \rangle$. Based on our previous theoretical analysis, one expects a scaling $\mathcal{N}_D \propto \langle c^{\dagger} c\rangle^{1/2}$ when the mean number of phonons increases. To test this idea, we work on the Hilbert space $\mathcal{H}_i$ (two-level system with states $|g\rangle$ and $|e\rangle$), and after solving the Heisenberg equation of motion for the effective Rabi model in the mean-field approximation, we found the following set of nonlinear equations 
\begin{eqnarray}
    \dot{\alpha} &=& -i\omega_{\rm ph} \alpha -i\left(g^{\ast}\langle \sigma_+\rangle + g\langle \sigma_-\rangle\right)-\gamma_{\rm ph} \alpha, \label{MF1}\\
    \langle \dot{\sigma}_+\rangle &=& i\omega_{\rm s} \langle \sigma_+ \rangle - ig\left(\alpha+\alpha^{\ast} \right)\langle \sigma_z \rangle - {\Gamma \over 2} \langle \sigma_+ \rangle, \label{MF2}\\
    \langle \dot{\sigma}_z\rangle &=&   2i\left(\alpha+\alpha^{\ast} \right)\left(g\langle \sigma_- \rangle - g^{\ast}\langle \sigma_+ \rangle\right) -\Gamma \langle \sigma_z \rangle. \label{MF3}
\end{eqnarray}

Here, $\Gamma$ is the effective decay rate of the populations. In Fig.~\ref{fig:Figure3}(a), we plot the behavior of the dynamical non-Markovianity in terms of the initial mean number of phonons using the mean field equations~\eqref{MF1}-\eqref{MF3}. As expected, the dynamical non-Markovianity exhibits the theoretically predicted scaling $\mathcal{N}_D = a \langle c^{\dagger}(0) c(0)\rangle^{1/2} = a|\alpha(0)|$. The scaling factor $a = a(T)$ depends on the temperature, and the expected behavior is shown in the inset of Fig.~\ref{fig:Figure3}. As expected, the factor $a(T)$ decreases with temperature, illustrating that thermal effects deteriorate the non-Markovian response of the point defect. The physical reason behind this thermal effect on the non-Markovianity is related to the damping 
rates, which increase with temperature, diminishing oscillations of the trace distance. \par

Until now, we have discussed the role of longitudinal fields, number of phonons, and coupling for the particular case of a magnetic field aligned with the symmetry axis of the SiV$^{-}$ center. In the following subsection, we shall discuss the four-level description of the SiV$^{-}$, leading to a more complex characterization of non-Markovianity.

\subsection{Four-level system: effect of transverse field and Breuer-Laine-Piilo measure of non-Markovianity} \label{FoureLevelSystem}

\begin{figure*}
\centering
\includegraphics[width = 0.8 \linewidth]{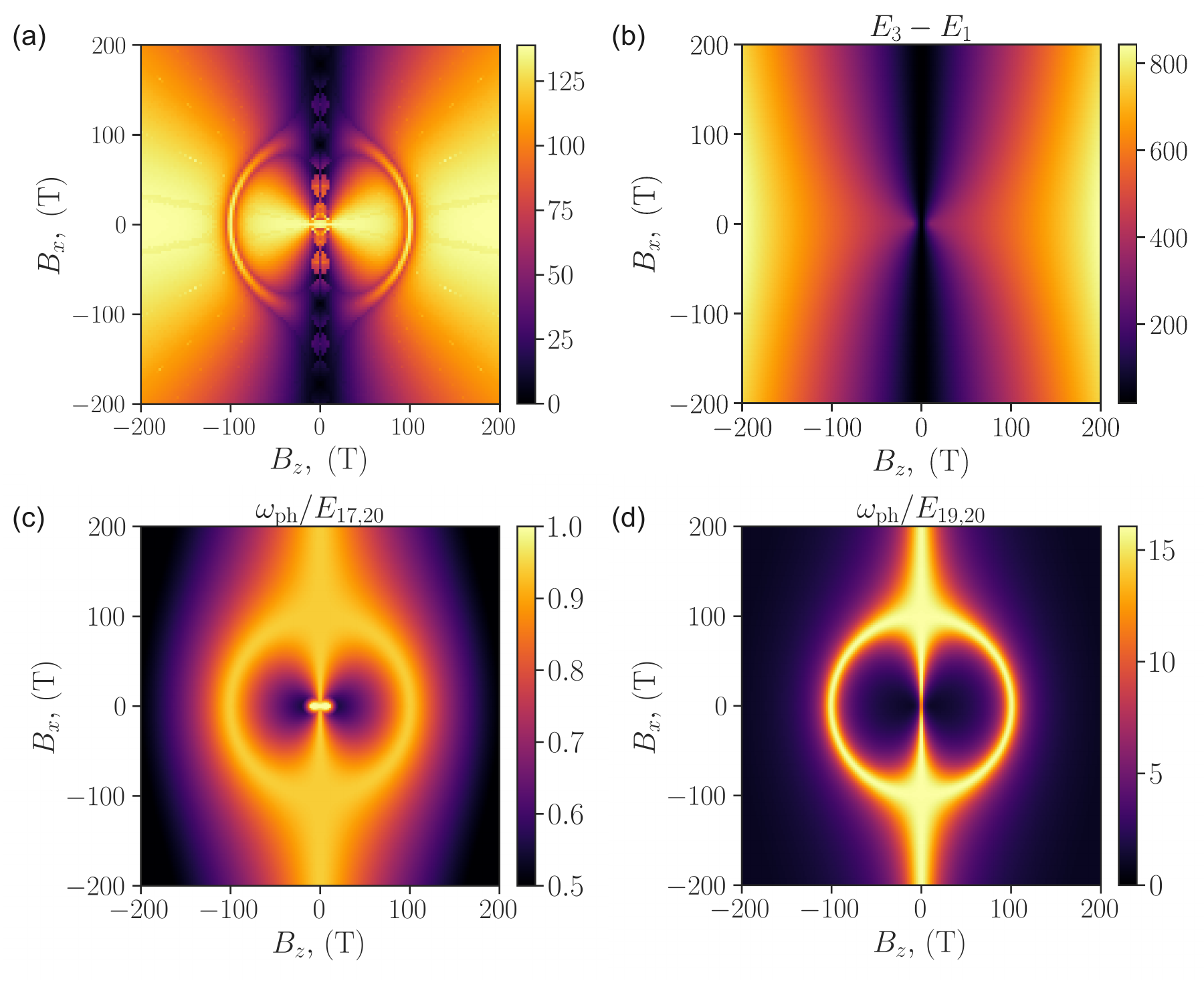}
\caption{(a) Pattern of the degree of non-Markovianity obtained from the BLP measure~\eqref{BLP} as a function of the longitudinal $B_z$ and transverse $B_x$ magnetic field components in the region $B_{x,z} \in [-200,200]$ T. (b) Energy gap $E_3-E_1$ (in GHz units), where $H_{\rm SiV}|i\rangle = E_i |i\rangle$ as function of the magnetic fields components. Ratio between the phonon mode $\omega_{\rm ph}=\Delta$ and energy gap $E_{i,j}$ for (c) $(i,j) = (17,20)$ and (d) $(i,j) = (19,20)$. For calculations of spectrum in (c) and (d), we use the full Hamiltonian given in Eq.~\eqref{SingleModeHamiltonian}.}
\label{fig:Figure4}
\end{figure*}

Now, we consider the effect of adding a transverse magnetic field on the non-Markovian dynamics. To do this, we consider $\mathbf{B} = (B_x,0,B_z)$ into the system Hamiltonian~\eqref{Hamiltonian}. In this case, the Hamiltonian of the system is described by ($\hbar = 1$)
\begin{eqnarray} \label{Htransversefield}
    H &=& \sum_{n=1}^{4}E_n |n\rangle \langle n| + {b_x \over 2}   \left(|1\rangle \langle 4| +|2\rangle \langle 3| + h.c.\right) +  \omega_{\rm ph}c^{\dagger} c\nonumber  \\ 
    & & + (c^{\dagger}+c)\left[g_1(L_-+L_+)-ig_2(L_- - L_+)\right], \nonumber \\
\end{eqnarray}
where $E_n$ are the energies introduced in Eqs.~\eqref{E1}-\eqref{E4} and $b_x=\gamma_s B_x$. We note that the transverse component $B_x$ induce transitions between states $|1\rangle \leftrightarrow |4\rangle$ and $|2\rangle \leftrightarrow |3\rangle$. These orbital preserving transitions (\textit{i.e.} without changing $|e_{\pm}\rangle$ states) are a consequence of the symmetries of the SiV$^{-}$ Hamiltonian since the orbital Zeeman effect only has the term $f \gamma_{\rm L} S_z L_z$ (reduced by a Ham factor $f \approx 0.1$). From a dynamical point of view, where a transverse field is present (along $x$ or $y$ directions), the SiV$^{-}$ must be modeled as a four-level system. We remark that the transverse field component $B_x$ can induce oscillations on the trace distance $D(t) = (1/2)||\rho_s(t)-\rho_{\rm SS}||$ defined on Eq.~\eqref{DNM}, which is not related to a non-Markovian behavior induced by phonons. Thus, to correctly quantify the phonon-induced non-Markovianity under the presence of the transverse field component, the Breuer-Laine-Piilo (BLP) measure must be used. The BLP measure is defined as~\cite {BLP2009, Breuer2010}
\begin{equation} \label{BLP}
    \mathcal{N}_{\rm BLP} = \max_{\rho_1(0),\,\rho_2(0)} \int_{\dot{D}>0}^{}\dot{D}(\rho_1(t),\rho_2(t)) dt,
\end{equation}
where $D(\rho_1(t),\rho_2(t)) = (1/2)||\rho_1(t)-\rho_2(t)||$ and $\rho_j(t)$ is the evolved density matrix for the initial state $\rho_j(0)$ ($j=1,2$) according to the Hamiltonian~\eqref{Htransversefield} and master equation~\eqref{MasterEquationSiVSingleMode}. The computational complexity of calculating $\mathcal{N}_{\rm BLP}$ is related to the calculation of the global maximum of a functional $D(\rho_1(t),\rho_2(t))$ that spans all possible initial states $\rho_i(0)$ ($i=1,2$) in the Hilbert space of the SiV$^{-}$ center. \par

In particular, to study non-Markovian effects, we focus on the following parametrization of uncorrelated initial states of the SiV$^{-}$ center density matrix
\begin{eqnarray} \label{IC}
    \rho(0) = \rho_{\rm orb}(0) \otimes \rho_{\rm spin}(0),
\end{eqnarray}
where we use pure states defined as $\rho_{X}(0) = |\Psi_{X}(0)\rangle \langle \Psi_{X}(0)|$ with $X \in  \{\text{orb},\text{spin}\}$. The initial states given above are motivated by the fact that initialization on the SiV$^{-}$ center can be realized using a magnetic field and mixing states $\ket{1} \leftrightarrow \ket{4}$ and $\ket{2} \leftrightarrow \ket{3}$ or by modifying the orbital states and mixing states $\ket{1} \leftrightarrow \ket{2}$ and $\ket{2} \leftrightarrow \ket{4}$. All these scenarios are described by the initial condition presented in Eq.~\label{IC}. \par

Using the Bloch representation $|\Psi_{\rm orb}(0)\rangle = \cos(\theta_1/2)|e_x\rangle + e^{i \phi_1}\sin(\theta_1/2)|g_y\rangle$ and $|\Psi_{\rm spin}(0)\rangle = \cos(\theta_2/2)\ket{\uparrow}+ e^{i \phi_2}\sin(\theta_2/2)\ket{\downarrow}$, we naturally introduce the set of parameters $\mathbf{x}_j=(\theta_1^j,\theta_2^j,\phi_1^j,\phi_2^j)$ required to write each initial state $\rho_j(0)$ ($j=1,2$). The problem of estimating the BLP measure, $\mathcal{N}_{\rm BLP}$, is translated into the following constrained nonlinear optimization problem 
\begin{eqnarray}
    &&\min_{\mathbf{x}} \mathcal{L}(\mathbf{x}), \quad \mathbf{x} = (\mathbf{x}_1,\mathbf{x}_2)  \nonumber \\
    \mathcal{L}(\mathbf{x}) &=&-\int_{\dot{D}>0}^{}D(\rho_1(t,\mathbf{x_1}),\rho_2(t,\mathbf{x_2})) dt,  \nonumber \\
    \mbox{subject to} && \; \mathbf{x}_{\rm lb}\leq \mathbf{x} \leq \mathbf{x}_{\rm ub}, 
\end{eqnarray}
where the physical constraint over the angles of the Bloch parametrization is equivalent to $\mathbf{x}_{\rm lb} = [0,0,0,0,0,0,0,0]$ (lower bound) and $\mathbf{x}_{\rm ub} = \pi[1,1,2,2,1,1,2,2]$ (upper bound). Here, the inequality $\mathbf{x}_{\rm lb}\leq \mathbf{x} \leq \mathbf{x}_{\rm ub}$ is the element-wise inequality over each component of the vector $\mathbf{x}= [x_1,x_2,...,x_8]$. For instance, $0\leq x_1 \leq \pi$ and $0\leq x_3 \leq 2\pi$. The functional $\mathcal{L}(\mathbf{x})$ must be calculated by propagating the initial conditions $\rho_j(0)$ according to the dynamical map described in Eq.~\eqref{MasterEquationSiVSingleMode}. In appendix~\ref{appendix:e}, we explain a robust computational implementation to find the global minimum of the functional $\mathcal{L}(\mathbf{x})$. \par

 We use the symmetry behavior $B_x \rightarrow -B_x$ and $B_z \rightarrow -B_z$, which is expected from a molecule structure with the inherent geometrical reflections symmetries $x \rightarrow -x$ and $z \rightarrow -z$. Therefore, our calculations of the degree of non-Markovianity were calculated for $B_x, B_z>0$, and then we applied the corresponding reflections. In Fig.~\ref{fig:Figure4}(a), we show the complex pattern of non-Markovianity by numerical computing the BLP measure in terms of the longitudinal and transverse magnetic field components. A detailed zoom on the range $-20 \; \mbox{T} \leq B_x,B_z \leq 20 \; \mbox{T}$ is presented in Appendix~\ref{appendix:e} (see Fig.~\eqref{fig:Figure7}). The pattern observed in Fig.~\ref{fig:Figure4}(a) shows a background behavior that can be explained by resonances on the SiV$^{-}$ energy levels and the ring feature for $(B_x+B_z)^{1/2} \approx 100$ T is evidence of a more complex resonance process.

\begin{figure*}
\centering
\includegraphics[width = 0.8 \linewidth]{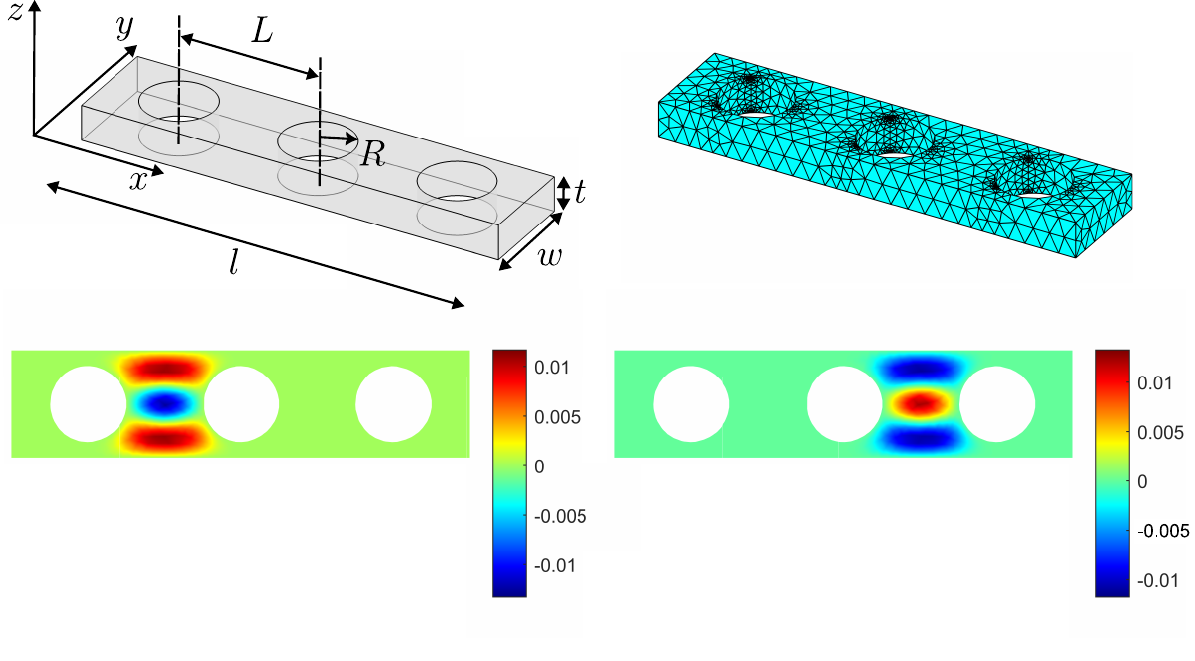}
\caption{Numerical calculation of localized phonon modes of frequencies $\omega_{\rm ph}/2\pi \approx 50$ GHz $= \Delta$ (SiV$^{-}$ energy gap at $\mathbf{B}=\mathbf{0}$) where circular holes are placed between a distance $L= 0.4 \; \mu$m in diamond structure with dimensions $(l, w, t) = (1.2, 0.28, 0.1) \mu$m. Colors indicate the value of the strain tensor in dimensionless units related to the e proportion of deformation $\Delta L/L_0$. Computational simulations were conducted using the PDE toolbox of MATLAB, where we used a mesh with quadratic tetrahedral elements. Localized vibrational modes with frequencies close to 50 GHz are depicted in the bottom panel (top view of the material), where oscillations are confined between holes.}
\label{fig:Figure5}
\end{figure*}

The background Fig.~\ref{fig:Figure4}(a) is intimately related to the behavior of the energy transitions $E_3-E_1$ (or $E_4-E_1$), which is calculated in Fig.~\ref{fig:Figure4}(b), in GHz units. As the magnetic field component $(B_x,B_z)$ varies, we note that the energy splitting response is similar to the background of the BLP measure. The latter is physically intuitive since, as discussed in the single phonon mode case, the degree of non-Markovianity is sensitive to energy resonances. However, the ring feature is explained by a more intriguing mechanism. To understand this ring behavior, we numerically analyze the energy spectrum of the Hamiltonian $H = H_{\rm SiV} + H_{\rm ph} + H_{\rm SiV-ph}$ given in Eq.~\eqref{SingleModeHamiltonian}, by including the effect of phonons. By solving the eigenvalue problem $H|\Phi_n\rangle = E_n|\Phi_n\rangle$ we can calculate the dimensionless factor $\omega_{\rm ph}/E_{n,m}$, where $E_{n,m}=E_n-E_m$. In Figs.~\ref{fig:Figure4}(c) and (d), we observe two cases where the pattern presents a ring feature around 100 T. However, only Fig.~\ref{fig:Figure4}(c) is relevant since there is a perfect resonance ($\omega_{\rm ph}/E_{17,20} \approx 1$ for $(B_x+B_z)^{1/2} \approx 100$ T) of the polaritonic states of Hamiltonian $H$ with the phonon mode. Besides the impossibility of finding an approximate analytical expression for the BLP measure, we believe these physical insights are relevant for this or other similar systems. \par

In this first part of the work, we focused on the non-Markovian behavior induced by a single phonon mode. In the next section, we will introduce the general model to incorporate a structured phonon environment at thermal equilibrium.

\section{Non-Markovian dynamics induced by a structured phonon environment} 

To introduce the effect of a structured phonon bath on the SiV$^{-}$ dynamics, we will begin with the strain interaction acting on the energy levels of the SiV$^{-}$~\cite{Kepesidis2016,Lemonde2018}. In the electronic basis spawned by the orbital states $\ket{e_{gx}}$ and $\ket{e_{gy}}$, the strain Hamiltonian can be written as~\cite{HeppThesis,Kepesidis2016}
\begin{equation}\label{Strain}
H_{\rm strain} = \left(\delta \mathds{1}_{\rm orb} + \alpha \sigma_z +\beta \sigma_x\right) \otimes \mathds{1}_s,
\end{equation}
where $\mathds{1}_{\rm orb} = |e_{gx}\rangle \langle e_{gx}|+|e_{gy}\rangle \langle e_{gy}|$ and $\mathds{1}_s = \ket {\uparrow}\!\bra{\uparrow} + \ket{\downarrow}\!\bra{\downarrow}$ are the orbital and spin identity matrices, respectively, $\delta = g_0(\gamma_{xx}+\gamma_{yy})$, $\alpha = g_1 (\gamma_{xx}-\gamma_{yy})$, and $\beta = g_2 \gamma_{xy}$. Here, $g_0$ and $g_1$ are coupling parameters associated with local lattice distortions around the defect, and $\hat{\gamma}_{ij} = (\partial \hat{u}_i/\partial x_j+\partial \hat{u}_j/\partial x_i)/2$ is the symmetrical strain tensor, with $\hat{u}_i$ representing quantized displacements along the Cartesian directions $x_1 = x$, $x_2 = y$, and $x_3 = z$~\cite{Kepesidis2016}. By decomposing the local displacement field using phonon operators, \textit{i.e.}, $\vec{\hat{u}} = \sum_n(\vec{u}_n c_n + \supervec{u}_n^{\ast}c_n^{\dagger})$, we obtain the following expression for the electron-phonon Hamiltonian 
\begin{equation} \label{e-ph-interaction}
H_{e-ph} =\sum_{n}\sum_{i,j}\left(\lambda_{ij,n} c_n |\varphi_i\rangle \langle \varphi_j| + \lambda_{ij,n}^{\ast} c_n^{\dagger} |\varphi_j\rangle \langle \varphi_i| \right),
\end{equation}
where $c_n$ ($c_{n}^{\dagger}$) is the annihilation (creation) phonon operators and $\lambda_{ij,n}$ are the electron-phonon coupling constants between different eigenstates of the SiV$^{-}$ center and vibrational modes. From elasticity theory we found that $\lambda_{ij,n} = A_{ij}g_n + B_{ij} f_n$, where $A_{ij} = \langle \varphi_i|(\sigma_z \otimes \mathds{1}_{s}) |\varphi_j\rangle$ and $B_{ij} = \langle \varphi_i| (\sigma_x \otimes \mathds{1}_{s}) |\varphi_j\rangle$ accounts for expectation values of the orbital operators $\sigma_i \otimes \mathds{1}_{s}$ ($i=x,z$) in the eigenbasis $\ket{\varphi_i}$. Magnetic field effects are hidden in the functional dependence of the eigenstates $\ket{\varphi}_i$ (see Eq.~\eqref{Hamiltonian}). Therefore, the presented electron-phonon Hamiltonian given in Eq.~\eqref{e-ph-interaction} is completely general, and for non-longitudinal magnetic fields, one must numerically compute $A_{ij}$ and $B_{ij}$. \par

On the other hand, the contribution of different phonon modes $n$ are encapsulated into the parameters $g_n = g_1\left(\partial_x \hat{u}_{n,x}-\partial_y \hat{u}_{n,y} \right)$ and $f_n = (1/2)g_2 \left(\partial_x \hat{u}_{n,y}+\partial_y \hat{u}_{n,x} \right)$, which depends on the vibrational profile $\hat{u}_{n,\alpha} = \vec{\hat{u}}_n \cdot \mathbf{e}_{\alpha}$. A detailed derivation of the electron-phonon Hamiltonian is presented in Appendix~\ref{appendix:a}. For the SiV$^{-}$, a generic environment can be modeled by introducing the following spectral density functions
\begin{eqnarray} 
J_1(\omega)=\sum_n |g_n|^2\delta(\omega-\omega_n), \label{J1}\\
J_2(\omega)=\sum_n |f_n|^2\delta(\omega-\omega_n).  \label{J2}
\end{eqnarray}

In principle, for a fixed set of experimental parameters, the open dynamics derived from the defect-phonon interaction is entirely determined by the shape of $J_1(\omega)$ and $J_2(\omega)$. However, phonons in a cantilever, waveguide, or phononic crystal can interact very differently with the localized orbital states of the defect; therefore, these nanodevices are characterized by a completely different spectral density function. This inherent geometry-dependent behavior of the phononic spectral density functions can be explored as a new tool for engineering and controlling quantum non-Markovianity in solid-state systems, which will be discussed in the next sections. The time evolution of the SiV$^{-}$ density matrix $\rho(t)$ follows a time-local master equation~($\hbar = 1$)
\begin{eqnarray}\label{MasterEquation}
\dot{\rho} &=& \sum_{i\neq j}\Gamma_{ij}(t)\left[\sigma_{ij}\rho\sigma_{ij}^{\dag}-\frac{1}{2}\{\sigma_{ij}^{\dag}\sigma_{ij},\rho\}\right] \nonumber \\
&& +\sum_{i,j}\Omega_{ij}(t)\left[\sigma_{ii}\rho\sigma_{jj}^{\dag}-\frac{1}{2}\{\sigma_{ii}^{\dag}\sigma_{jj},\rho\}\right],
\end{eqnarray}
where $\rho(t) = \sum_{i,j} \rho_{ij}(t) |\varphi_i\rangle \langle \varphi_j|$ is written in the interaction picture, $\sigma_{ij} = |\varphi_i\rangle \langle \varphi_j|$, with $|\varphi_i\rangle$ being the eigenstates of the SiV$^{-}$ center. The first and second terms of the right-hand of Eq.~\eqref{MasterEquation} are recognized as the amplitude damping (energy-exchange) and pure-dephasing noises, respectively, derived from the electron-phonon coupling. For further details of the derivation of the master equation and the expression of time-dependent rates $\Gamma_{ij}(t)$ and $\Omega_{ij}(t)$, see Appendix~\ref{appendix:b}. Note that the time-dependent rates $\Gamma_{ij}(t)$ and $\Omega_{ij}(t)$, which are defined in Eqs.~\eqref{GammaRate}-\eqref{OmegaRate}, are the source of non-Markovianity in this model. These rates depend on the spectral density functions $J_1(\omega)$ and $J_2(\omega)$, where the role of temperature is encapsulated by the mean number of phonons at thermal equilibrium. \par

Now, we will use the time-dependent master equation given in Eq.~\eqref{MasterEquation} to discuss the case of a structured phonon bath with a dominant interaction with a localized phonon mode and the role of temperature.

\subsection{Geometrical confinement of phonon modes}

Recent theoretical studies in phononic crystals and color centers have implemented numerical calculations using elasticity theory to model mechanical compression modes in one-dimensional diamond structures~\cite{Nori2019}. Similar methods are found in two- and three-dimensional diamond nanostructures~\cite{Lemonde2018, Wang2018}. In the present project, we will calculate the mechanical properties in the framework of the elasticity theory~\cite{ref59}. To this end, we perform numerical calculations to find the phonon displacement profile using a phenomenological approach based on a continuous method. Mechanical modes at the position $\mathbf{r}$ and time $t$ are described by the continuum field $\mathbf{Q}(\mathbf{r},t)$, where the following equation of motion is satisfied~\cite{Lemonde2018, Nori2019, Wang2018}:
\begin{equation} \label{QModes}
    \rho\frac{\partial^2}{\partial t^2} \mathbf{Q}(\mathbf{r},t) = (\lambda_0 + \mu)\nabla (\nabla \cdot \mathbf{Q}(\mathbf{r},t)) + \mu \nabla^2 \mathbf{Q}(\mathbf{r},t),
\end{equation}
where $\rho = 2329 \, \text{kg/m}^3$ is the density of diamond, and $(\lambda_0, \mu)$ are the Lam\'e constants given by the relations $\lambda_0 = \nu E/[(1 + \nu)(1 - 2\nu)]$ and $\mu = E/[2(1 + \nu)]$, with $\nu = 0.2$ and $E = 1050 \, \text{GPa}$ being the Poisson ratio and Young’s modulus in diamond, respectively~\cite{Nori2019}. The numerical calculation of Eq.~\eqref{QModes} is performed using MATLAB. \par

To apply Eq.~\eqref{QModes} and find vibrational profiles of the phonon modes, we focus on a finite-sized quasi-one-dimensional phononic crystal with circular holes, as illustrated in Fig.~\ref{fig:Figure5}. This is motivated by Ref.~\cite{Nori2019}, where elliptical holes in a diamond structure are numerically modeled to enhance and control band-gap-engineered spin-phonon interactions. In our work, we focus on the possibility of having localized vibrations between two circular holes with frequencies close to the SiV$^{-}$ energy gap $\Delta \approx 2\pi \times 50$ GHz. A vibrational mode confined in space resembles the original Jaynes-Cummings model, where a single photonic mode can interact with an atom-like system. However, under a realistic scenario, a continuum of phonon modes close to the SiV$^{-}$ energy gap will contribute to the dynamics. Therefore, we introduce the following phenomenological spectral density function
\begin{equation} \label{SDF}
    J(\omega) = J_0 {\omega^3 \over (\omega / \Delta)^2 + 1}  {(\Gamma/2) \over (\omega-\Delta)^2+(\Gamma/2)^2},    
\end{equation}
where $J_0$ is the amplitude and $\Gamma$ is the width. This spectral density function has been used to model phonon interactions in quantum dots~\cite{Rae2002}
, color center~\cite{Ariel2020,Ariel2016}, and to study non-Markovian effects in spin systems coupled to phonons~\cite{Norambuena2023}. In particular, the spectral density functions must satisfied $\int J_1(\omega)d\omega = \sum_n |g_n|^2$ and $\int J_2(\omega)d\omega = \sum_n |f_n|^2$. For the simulation, we use $J_0$ parameters such that $\sum_n |g_n|^2 =\sum_n |f_n|^2 \sim (1-10)\Delta^2$, which is a moderate Huang-Rhys factor ($1-10$) for a resonant phonon mode with the SiV$^{-}$ energy gap at zero magnetic field. \par

The physical intuition behind Eq.~\eqref{SDF} is the acoustic contribution at low frequencies $J(\omega) \propto \omega^3$and the strong phonon interaction with a quasi-localized phonon mode at $\omega=\Delta$. These considerations lead to a phenomenological description of the non-Markovian dynamics for structured phonon baths. Thus, to numerically evaluate the BLP measure, we will implement $J_1(\omega)=J_2(\omega)=J(\omega)$ into the time-local master equation shown in Eq.~\eqref{MasterEquation}. We apply the same Bloch parametrization of the initial states of the SiV$^{-}$ center, as explained in Sec.~\ref{FoureLevelSystem}.  \par

\begin{figure}[h!]
\centering
\includegraphics[width = 1 \linewidth]{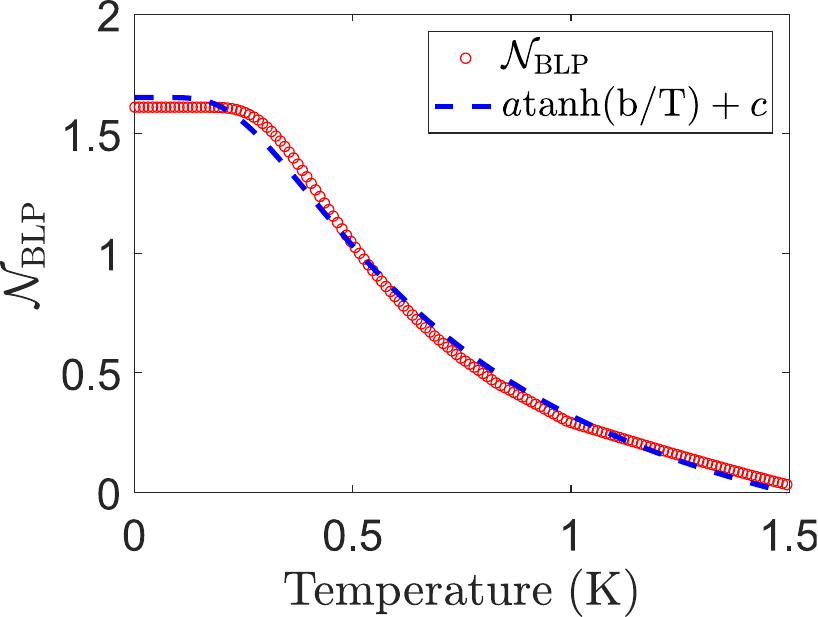}
\caption{Degree of non-Markovianity as a function of temperature for a structured phonon bath model with Lorentzian-like spectral density function with $J_0 = 4.55 \Delta$. The fit model is defined by $a = 2.386 \pm 0.016 $, $b=0.475 \pm 0.006$, and $c = -0.734 \pm 0.017$ with a mean square error equal to $1.4 \times 10^{-3}$.}
\label{fig:Figure6}
\end{figure}

\subsection{Degree of non-Markovianity and role of temperature}
In this section, we discuss the role of temperature on the degree of non-Markovianity. In our model, temperature arises from the statistical description of phonons according to the Bose-Einstein distribution of the mean number of phonons $N(\omega, T)=[\mbox{exp}(\hbar\omega/(k_BT))-1]^{-1}$, where $k_B$ is the Boltzmann factor, $\omega$ is the phonon frequency, and $T$ is the temperature. This thermal occupation factor appears in the definition of the function $\Gamma_{ij}(t)$ and $\Omega_{ij}(t)$ in Eq.~\eqref{MasterEquation}, see Appendix~\ref{appendix:b}. In particular, all time-dependent damping rates $\gamma_{ij}^{\omega}(t)$ shown in Eqs.~\eqref{g1}-\eqref{g8} depends on the factor $n(\omega)$, which is usual in the time-local Lindblad formalism for open quantum systems.

Intuitively, a narrow peak in the spectral density function $J_i(\omega)$ at $\omega \approx \Delta$ will select phonons around the SiV$^{-}$ energy gap $\Delta$ to contribute to the time-dependent rates. However, the amplitude of the rates can be drastically modified by the thermal factors at the corresponding resonant $\omega \approx \Delta$. The higher the temperature, the longer the effect of temperature on increasing the rates, which deteriorates the non-markovian response of the system. Thus, the BLP measure of non-Markovianity is expected to decrease with temperature or, equivalently, an inverse relation with damping rates amplitude. The critical question is, what is the thermal window to observe some non-Markovian effects? To answer this, in Fig.~\ref {fig:Figure6}, we numerically calculate the degree of non-Markovianity as a function of temperature using the phenomenological spectral density function introduced in Eq.~\eqref{SDF} but using a simulated annealing method for the optimization problem. Surprisingly, the BLP measure shows non-Markovian features in the SiV$^{-}$ center only appear for temperatures below $1.5$ K. In particular, we note that absorption and emission amplitudes scales as $\gamma \sim  \gamma_0[2N(\omega,T)+1] = \mbox{tanh}[\hbar \omega / (k_B T)]$, and therefore, the expected inverse relation $\mathcal{N}_{\rm BLP} \propto \gamma^{-1} \propto \mbox{tanh}[\hbar \omega / (k_B T)]$. The latter motivates the fitting curve $\mathcal{N}_{\rm BLP}=a\mbox{tanh}(b/T)+c$, where $a,b,c$ are fitting parameters. In Fig.~\ref {fig:Figure6}, we plot this fitting model (dashed lines) to observe the good agreement with numerical calculations of the degree of non-Markovianity. \par

One possible strategy to enhance the non-Markovian behavior for more extended temperatures could be to engineer short-range interactions between different SiV centers. This could lead to a quantum simulation of the Jaynes-Cummings-Hubbard (JCH) model, where cooperative interactions between adjacent SiV centers could improve memory effects. 

\section{Conclusions}
In this work, we have presented a detailed and physically motivated study of the non-Markovian behavior of the SiV$^{-}$ center in diamond for single mode and phonon bath models. First, we introduce a formal Hamiltonian description of the open dynamics of a single mode coupled to the energy levels of a SiV center. Using theoretical and numerical calculations, we study the role of the magnetic field $B_z$ and $B_x$ for initial Fock states. We showed that for a longitudinal magnetic field, $\mathbf{B}= (0,0,B_z)$ the dynamics can be represented by an effective Rabi model, where resonant conditions $\omega_{\rm ph} = \Delta$ leads to a maximum dynamical non-Markovianity ($\mathcal{N}_D$). We found approximate analytical expressions to understand our simulations. However, by including a transverse component and suing $\mathbf{B}= (B_x,0, B_z)$, we found exotic non-Markovian patterns in the $B_x$-$B_z$ domain. Also, we illustrate that a mean-field analysis sheds light on the effect of initialization of the system with a high number of phonons. We found a scaling $\mathcal{N}_D \propto \langle c^{\dagger} c\rangle^{1/2}$ consistent with our theoretical calculation. \par

In the case of a structured phonon bath, we focus the discussion on the case of having a localized phonon in resonance with the energy gap of the SiV$^{-}$ center at zero magnetic field. We motivate the existence of such localized vibrations by simulation a continuous vibrational field in a finite-sized diamond structure with circular holes. Our computations simulations showed that there are geometrical configurations that allow the presence of localized vibrations between holes resonating with the SiV center. These calculations motivated the theoretical calculation of a time-local master equation for a phonon bath described by a phenomenological spectral density function with a Lorentzian shape. Under these conditions, we found that the BLP measure of non-Markovianity exists at temperatures below 1 K. We believe that this detailed study of non-Markovianity in a solid-state spin device can motivate the search for a novel phonon-based platform to enhance and eventually use non-Markovian effects into quantum technologies. 

\section{Acknowledgments}
We gratefully acknowledge Hossein Dinani and Francisco Albarr\'an-Arriagada for their stimulating conversations. A.N. acknowledges the financial support from the project Fondecyt Iniciaci\'on No. 11220266. D.T. and D.C. acknowledge support from the Universidad Mayor through the Doctoral fellowship. V.C. acknowledges the high-performance computational resources
NCSA's Illinois Campus Cluster supercomputer provided under the Illinois Computes program, University of Illinois Urbana-Champaign. V.C. acknowledges the internship at the Multidisciplinary Center for Physics at the Universidad Mayor.
\appendix

\section{Derivation of the electron-phonon Hamiltonian}\label{appendix:a}

Local distortions near a point defect generate a modification of the electronic distribution of the defect via electron-ion interaction. To first order in the ion displacements and in the Born-Oppenheimer approximation, this local distortion effect can be modeled by the strain Hamiltonian~\cite{Kepesidis2016} 
\begin{equation}
H_{\mbox{\scriptsize strain}} = \sum_{i,j,\alpha,\beta}|\alpha \rangle \langle \alpha| V_{ij} |\beta \rangle \langle \beta| \hat{\gamma}_{ij},
\end{equation}
where $|\alpha\rangle$ is the electronic basis, $V_{ij}$ are couplings that involve the electron-ion Coulomb interaction~\cite{Maze2011}, and $\hat{\gamma}_{ij} = (\partial \hat{u}_i/\partial x_j+\partial \hat{u}_j/\partial x_i)/2$ is the strain tensor. For the SiV$^{-}$, the strain tensor can be expressed using matrices that transform according to the irreducible representations of the $C_{3v+i}$ group. This mathematical decomposition can be achieved by projecting the strain matrix in each irreducible representation by applying the following formula~\cite{Tinkham2003, Maze2011}
\begin{eqnarray}
\hat{\gamma}_{r} &=& {l_r \over h} \sum_{g \in C_{3v+i}} \chi^{r} (g)R^{\dagger} (g)\;  \hat{\gamma} R(g),  \nonumber \\
&& \hat{\gamma} = \left(\begin{array}{ccc}
                              \hat{\gamma}_{xx} & \hat{\gamma}_{xy} &  \hat{\gamma}_{xz} \\
                              \hat{\gamma}_{xy} & \hat{\gamma}_{yy} &  \hat{\gamma}_{yz} \\
                              \hat{\gamma}_{xz} & \hat{\gamma}_{yz} &  \hat{\gamma}_{zz} \\
                           \end{array} \right), \label{Project Strain}
\end{eqnarray}
where $\hat{\gamma}_{r}$ is the projection of the strain tensor into the irreducible representation $r$, $l_r$ is the dimension of the $r$-th irreducible representation, $h$ is the order of the group, $\chi^{r}(g)$ is the character of the group element $g$ in the irreducible representation $r$, and $R(g)$ is the matrix representation of the group element $g$ (see Character Table~\eqref{CharacterTableSiV}).
 
\begin{table}[]
\centering
\caption{Character table of the group $C_{3v+i}$. The first column shows the irreducible representations of the $C_{3v+i}$ group ($A_{1g}, A_{2g}, E_g, A_{1u}, A_{2u}, E_u$). The indices $g$ and $u$ denote the German words “gerade” (symmetric) and “ungerade” (antisymmetric) representations. The first row lists the symmetry operations of the group $C_{3v}$, organized by the classes of the group ($E, 2C_3, 3C_2, i, 2S_6, 3 \sigma$).}
\label{CharacterTableSiV}
\scalebox{1.1}{
\begin{tabular}{|c|r|r|r|r|r|r|}
\hline
 & $E$  & $2C_3$ & $3C_2$  & $i$  & $2S_6$  & $3\sigma$  \\ \hline
 $A_{1g}$ & 1 & 1 & 1 & 1 & 1 & 1  \\ \hline
 $A_{2g}$ & 1 & 1 & -1 & 1 & 1 & -1  \\ \hline
 $E_{g}$  & 2 & -1 & 0 & 2 & -1 & 0  \\ \hline
 $A_{1u}$ & 1 & 1 & 1 & -1 & -1 & -1  \\ \hline
 $A_{2u}$ & 1 & 1 & -1 & -1 & -1 & 1  \\ \hline
 $E_{u}$  & 2 & -1 & 0 & -2 & 1 & 0  \\ \hline
\end{tabular}}
\end{table}

Using the formula given in Eq.~\eqref{Project Strain}, we find 
\begin{equation}
\hat{\gamma} = \hat{\gamma}_{A_{1g}} + \hat{\gamma}_{E_g}, 
\end{equation}
where
\begin{eqnarray}
\hat{\gamma}_{A_{1g}} &=& \left[\begin{array}{ccc} 
                                      {1 \over 2}\left(\gamma_{xx} + \gamma_{yy} \right) & 0 & 0 \\
                                     0 & {1 \over 2}\left(\gamma_{xx} + \gamma_{yy} \right) & 0 \\
                                     0 & 0 & \gamma_{zz} \\
                                      \end{array} \right],\\                                      
\hat{\gamma}_{E_{g}} &=& \left[\begin{array}{ccc} 
                                      {1 \over 2}\left(\gamma_{xx} - \gamma_{yy} \right) & \gamma_{xy} & \gamma_{xz} \\
                                     \gamma_{xy} & {1 \over 2}\left(\gamma_{yy} - \gamma_{xx} \right) & \gamma_{yz} \\
                                    \gamma_{xz} & \gamma_{yz} & 0 \\
                                       \end{array} \right].
\end{eqnarray}

Due to the inversion symmetry of the SiV$^{-}$ center, the orbital degrees of freedom of the states within the ground and excited subspaces are characterized by parity~\cite{Tinkham2003}. Consequently, the expectation values of the form $\langle \alpha |V_{iz} | \alpha \rangle$ ($i=x,y,z$) vanish in both ground and excited subspaces. Therefore, in the electronic basis spawned by $\{\ket{e_{gx}}, \ket{e_{gy}}\}$, the strain Hamiltonian can be written as given in Eq.~\eqref{Strain}. Let us consider an arbitrary magnetic field $\vec{B}$. In such a case, the eigenvalues and eigenvectors of the SiV$^{-}$ are given by the equation $H|\varphi_i\rangle = E_i |\varphi_i\rangle$. In the basis $|\varphi_i\rangle$ the strain Hamiltonian can be written as
\begin{equation}
H_{\rm strain} = \delta \sum_{i}|i\rangle\langle i|+  \sum_{i,j}\left( \alpha A_{ij}+\beta B_{ij}\right)|i\rangle\langle j|,
\end{equation}
where $A_{ij} = \langle i|(\sigma_z \otimes \mathds{1}_{s}) |j\rangle$ and $B_{ij} = \langle i| (\sigma_x \otimes \mathds{1}_{s}) |j\rangle$, with $\mathds{1}_s = \ket{\uparrow}\!\bra{\uparrow} +\ket{\downarrow}\!\bra{\downarrow}$. The first term in the above Hamiltonian is a constant shift induced by symmetric distortions of the lattice. By decomposing the local displacement field as $\vec{\hat{u}} = \sum_n(\vec{u}_n \hat{c}_n + \supervec{u}_n^{\ast}\hat{c}_n^{\dagger})$, we recover the electron-phonon Hamiltonian given in Eq.~\eqref{e-ph-interaction}.

\section{Microscopic derivation of the time-local master equation}\label{appendix:b}

Assuming a phononic bath in thermal equilibrium and neglecting the initial correlations, we start our microscopic derivation by using the following master equation for the SiV$^{-}$ center in the interaction picture~\cite{Breuerbook}
\begin{equation}\label{FirstME}
\dot{\rho}= - \int_{0}^{t} dt' \mbox{Tr}_{\rm ph} \left(\left[H_I(t),\left[H_I(t'), \rho(t) \otimes \rho_{\rm ph} \right]\right]\right),
\end{equation}
where $\rho_{\rm ph} = \mbox{exp}(-\beta H_{\rm ph})/\mbox{Tr}_{\rm ph}
(\mbox{exp}(-\beta H_{\rm ph}))$ is the phononic density matrix, with $H_{\rm ph} = \sum_n \omega_n c_n^{\dagger}c_n$ being the phonon Hamiltonian. Here, $\rho(t)$ is the SiV$^{-}$ center density matrix in the interaction picture. On Eq.~\eqref{FirstME}, we have assumed the weak-coupling approximation, which is satisfied for structured phonon environments in diamond. In the interaction picture, the electron-phonon Hamiltonian can be written as
\begin{equation} \label{Hi(t)}
H_I(t) = \sum_{i=1}^4 S_{\alpha}(t) \otimes E_{\alpha}(t),
\end{equation}
where $S_{\alpha}(t) = \mbox{exp}(iH_{SiV}t)S_{\alpha}\mbox{exp}(-iH_{SiV}t)$ and $E_{\alpha}(t) = \mbox{exp}(iH_{\rm ph}t)E_{\alpha}\mbox{exp}(-iH_{\rm ph}t)$, with
\begin{eqnarray}
S_1 &=& \sum_{i,j}A_{ij}|\varphi_i\rangle \langle \varphi_j| = S_3^{\dagger}, \label{S1} \\
S_2 &=& \sum_{i,j}B_{ij}|\varphi_i\rangle \langle \varphi_j| = S_4^{\dagger}, \\
E_1 &=& \sum_n g_n C_n = E_3^{\dagger}, \\
E_4 &=& \sum_n f_n C_n = E_2^{\dagger}, \label{E4}
\end{eqnarray}

By inserting Eq.~\eqref{Hi(t)} into Eq.~\eqref{FirstME}, we obtain 
\begin{eqnarray}
 \dot{\rho} &=& \sum_{\omega,\alpha,\beta}\gamma_{\alpha,\beta}^{\omega}(t)\left[S_{\beta}(\omega)\rho S_{\alpha}^{\dag}(\omega)- \vphantom{\frac{1}{2}}\right. \nonumber \\
 &&\left. \frac{1}{2}\{S_{\alpha}^{\dag}(\omega)S_{\beta}(\omega),\rho\}\right],
\end{eqnarray}
where the operators $S_{\alpha}(\omega)$ are defined as
\begin{equation}\label{SpectralDecomposition}
S_{\alpha}(\omega) = \sum_{i,j}\delta(\omega_{ji}-\omega)|\varphi_i\rangle \langle \varphi_i| S_{\alpha} |\varphi_j\rangle \langle \varphi_j|,
\end{equation}
where $\omega_{ij} = (E_i-E_j)/\hbar$, $E_{i,j}$ are the eigenvalues of the SiV Hamiltonian, and $\delta(\omega_{ba}-\omega)$ is a Kronecker function, \textit{i.e.}, $\delta(x)= 1 $ for $x=0$ and $\delta(x)=0$ otherwise. By solving the spectral decomposition~\eqref{SpectralDecomposition}, where $\delta(\omega_{ba}-\omega)$ is a Kronecker function, \textit{i.e.}, $\delta(x)= 1 $ for $x=0$ and $\delta(x)=0$ otherwise. From Eq.~\eqref{SpectralDecomposition} we recover the time-local master equation given in Eq.~\eqref{MasterEquation}, where
\begin{eqnarray} 
\Gamma_{ij}(t) &=&  |A_{ij}|^2(\gamma_{11}^{\omega_{ji}}+\gamma_{33}^{\omega_{ji}})+|B_{ij}|^2(\gamma_{22}^{\omega_{ji}}+\gamma_{44}^{\omega_{ji}}) \\
&& +B_{ij}A_{ij}^*(\gamma_{12}^{\omega_{ji}}+\gamma_{34}^{\omega_{ji}})+B_{ij}^*A_{ji}(\gamma_{21}^{\omega_{ji}}+\gamma_{43}^{\omega_{ji}}), \nonumber  \label{GammaRate} \\
\\
\Omega_{ij}(t) &=& A_{ii}A_{jj}^*(\gamma_{11}^{\omega=0}+\gamma_{33}^{\omega = 0})+B_{ii}B_{jj}^*(\gamma_{22}^{\omega=0}+\gamma_{44}^{\omega=0}) \nonumber \\ 
&& +B_{ii}A_{jj}^*(\gamma_{12}^{\omega=0}+\gamma_{34}^{\omega=0})+B_{jj}A_{ii}^*(\gamma_{21}^{\omega=0}+\gamma_{43}^{\omega=0}).\nonumber 
\label{OmegaRate} \\
\end{eqnarray}
Here, the individual time-dependent rates are defined as
\begin{equation}
\gamma_{11}^{\omega}(t) = 2\int_0^{\infty}J_1(\omega')n(\omega')\frac{\sin(\omega+\omega')t}{\omega+\omega'}d\omega', \label{g1}
\end{equation}
\begin{equation}
\gamma_{12}^{\omega}(t) = 2\int_0^{\infty}J_3(\omega')n(\omega')\frac{\sin(\omega+\omega')t}{\omega+\omega'}d\omega', \label{g2}
\end{equation}
\begin{equation}
\gamma_{22}^{\omega}(t) = 2\int_0^{\infty}J_2(\omega')n(\omega')\frac{\sin(\omega+\omega')t}{\omega+\omega'}d\omega', \label{g3}
\end{equation}
\begin{equation}
\gamma_{21}^{\omega}(t) = 2\int_0^{\infty}J_3^{*}(\omega')n(\omega')\frac{\sin(\omega+\omega')t}{\omega+\omega'}d\omega', \label{g4}
\end{equation}
\begin{equation}
\gamma_{33}^{\omega}(t)=2\int_0^{\infty}J_{1}(\omega')(n(\omega')+1)\frac{\sin(\omega-\omega')t}{\omega-\omega'}d\omega', \label{g5}
\end{equation}
\begin{equation}
\gamma_{34}^{\omega}(t)=2\int_0^{\infty}J_{3}^{*}(\omega')(n(\omega')+1)\frac{\sin(\omega-\omega')t}{\omega-\omega'}d\omega', \label{g6}
\end{equation}
\begin{equation}
\gamma_{44}^{\omega}(t)=2\int_0^{\infty}J_{2}(\omega')(n(\omega')+1)\frac{\sin(\omega-\omega')t}{\omega-\omega'}d\omega', \label{g7}
\end{equation}
\begin{equation}
\gamma_{43}^{\omega}(t)=2\int_0^{\infty}J_{3}(\omega')(n(\omega')+1)\frac{\sin(\omega-\omega')t}{\omega-\omega'}d\omega', \label{g8}
\end{equation}
where $J_1(\omega)$ and $J_2(\omega)$ are the spectral density function introduced in Eqs.~\eqref{J1} and \eqref{J2}, and $J_3(\omega) = \sum_n g_n^{*}f_n \delta(\omega-\omega_n)$. 

\section{Approximate solution dynamical degree of non-Markovianity}\label{appendix:c}

The time derivative of the trace distance defined in Eq.~\eqref{DNM} is given by $\dot{D} = \mbox{sgn}(f(t))[ \dot{f} - \Gamma_0 f(t)]e^{-\Gamma_0 t}$, and non-Markovianity is manifested in the time intervals such that $\dot{D} > 0$. Here, we are using the sign function $\mbox{sgn}(x) = x /|x|$ and the property $|f| = \mbox{sgn}(f)f$. In our model, the trace distance always increases from $D(t_n^{\rm zeros})=0$ to some maximum value such that $\dot{D}(t_{n}^{\rm max})=0$, with $t_{n}^{\rm max}>t_n^{\rm zeros}$. Such times can be easily obtained by imposing the condition $|f(t_n^{\rm zeros})|=0$ and $\mbox{sgn}(f(t_{n}^{\rm max})) \dot{f}(t_{n}^{\rm max}) - \Gamma_0 |f(t_{n}^{\rm max})|=0$. We obtain
\begin{eqnarray}
    t_n^{\rm max} &=& \mu^{-1}\left(n \pi-\pi/2+\alpha_1\right), \\
    t_n^{\rm zeros}&=& \mu^{-1}\left(n \pi-\pi/2+\alpha_2\right),
\end{eqnarray}
where $n \in \mathbb{N}^+$, $\alpha_1 = \mbox{tan}^{-1}(C_2/C_1)$, and $\alpha_2 = \pi-\mbox{tan}^{-1}(|[\mu C-1+\Gamma_0 C_2]/[\mu C_2-\Gamma_0 C_1]|)$. From our numerical simulations we confirm that $\dot{D}>0$ in the time intervals $t \in [t_n^{\rm zeros},t_n^{\rm max}]$. Thus, using the fact that $D(t_n^{\rm zeros})=0$, we can recover the expression given in Eq.~\eqref{AnalyticDyNM}. We have used the approximation $|f(t_{n}^{\rm max})| \approx 1/2$, which has been corroborated using our simulations.

\section{Computational implementation to calculate the Breuer-Laine-Piilo measure}\label{appendix:e}

By optimizing over initial states $\rho_1(0)$ and $\rho_2(0)$, we calculate the BLP measure, $\mathcal{N}_{\rm BLP}$ (see Eq.~\eqref{BLP}), for different combinations of magnetic fields $(B_x,B_z)$. In our case, we determine the $\mathcal{N}_{\rm BLP}$ for $71^2 = 5041$ combinations of $71$ different $B_x$ and $B_z$ values, both ranging between $0\,\rm T$ and $200\, \rm T$. The data was then reflected about both the $B_x$ and $B_z$ axes, thus obtaining the $\mathcal{N}_{\rm BLP}$ for combinations in the entire $-200\,{\rm T}\leq B_x, B_z \leq 200\,{\rm T}$ domain. This reflection is justified by the symmetry of the SiV$^{-}$ center along the $x$- and $z$-axes. With this, we created the heat map shown in Fig.~\ref{fig:Figure4}(a), with $(71+70)^2=19881$ data points, each representing the $\mathcal{N}_{\rm BLP}$ for a given magnetic field $\mathbf{B}=(B_x,0,B_z)$. \par

\begin{figure}
\centering
\includegraphics[width = 1 \linewidth]{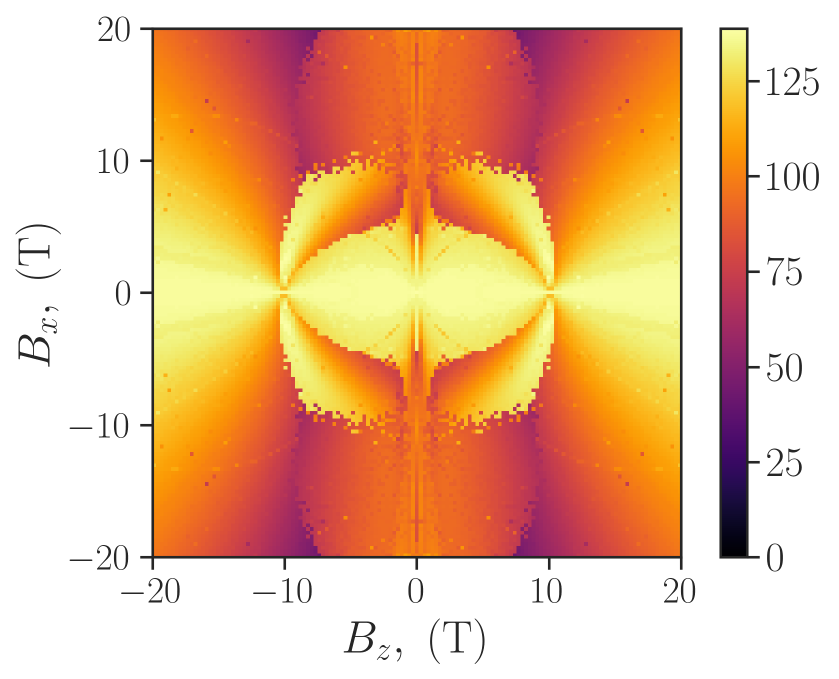}
\caption{Pattern of the degree of non-Markovianity obtained from the BLP measure~\eqref{BLP} as a function of the longitudinal $B_z$ and transverse $B_x$ magnetic field components in the region $B_{x,z} \in [-20,20]$ T. }
\label{fig:Figure7}
\end{figure}

To set up our calculations, we utilized QuTiP, an open-source Python library designed for simulating open quantum systems~\cite{qutip1,qutip2}. QuTiP enabled us to model our system and solve its dynamics efficiently. Then, to perform the optimization over our initial states $\rho_1(0)$ and $\rho_2(0)$, we utilized Python SciPy's Differential Evolution (DE) optimization method. We used the DE method since it is a stochastic population-based algorithm used for global optimization problems, and it is fast enough to estimate the degree of non-Markovianity. \par

However, optimizing over solutions to the dynamics of a four-level system is computationally expensive, and our initial tests indicated that optimizing over eight parameters on $5041$ different points would have prohibitively long execution times. To address this issue, we developed a mixed-resolution optimization approach. Our technique involves initially performing the optimization with a lower resolution, characterized by a shorter time range of integration $t_f - t_0$ and a lower point density on the $\mathcal{N}_D$ curve. Once optimal parameters are identified, we then compute the $\mathcal{N}_D$ with a higher resolution, which entails a longer time range of integration $t_f - t_0$, and a higher point density on the $\mathcal{N}_D$ curve to provide a more accurate result. As execution time grows linearly with the number of points, this method significantly improves efficiency. \par

While this mixed-resolution optimization approach effectively reduces computational load and execution time, its validity depends critically on having a sufficient number of points in the low-resolution stage; plotting an insufficient number of points could lead to sub-optimal results as such insufficiency could cause the optimizer to overlook points that are essential to properly characterizing the $\mathcal{N}_D$ curve. Hence, empirical testing is essential to determine the optimal number of points that balance result accuracy and validity with computational efficiency. In our case, we plotted $10^5$ points over the interval $(|g|t_0\!=\!0,\,|g|t_f\!=\!30)$ for the low-resolution stage, and $10^6$ points over the interval $(|g|t_0\!=\!0,\,|g|t_f\!=\!100)$ in the high-resolution stage. This approach reduced the execution time by a factor of $10$. \par

To further accelerate execution, we used Joblib, a Python library that facilitates parallel processing. This technique enables the simultaneous execution of multiple computationally demanding operations across multiple $N$ CPU cores, effectively reducing the total computation time by a factor of $N$. To fully leverage this parallelization, the simulations were executed on NCSA's Illinois Campus Cluster supercomputer under the Illinois Computes program, simultaneously utilizing $40$ nodes with $128$ CPU cores each (\textit{i.e.}, a total of $5120$ CPU cores) for parallel processing. Combining our mixed-resolution optimization approach with parallel processing across $5120$ CPU cores reduced execution time by a factor of $51200$. The total computation time for the optimization process, which involved $5041$ points, was approximately 36 hours.

Fig.~\ref{fig:Figure7} was generated by repeating the same process, calculating the same number of data points, but restricted to the  $-20\,{\rm T}\leq B_x, B_z \leq 20\,{\rm T}$ domain to reveal greater detail.

\section{Calculation of the phonon profile using MATLAB}\label{appendix:f}

Calculating the phonon profile of Eq.~\eqref{QModes} requires a modal analysis by proposing a solution of the form $\mathbf{Q}(\mathbf{r},t) = \mathbf{U}(\mathbf{r})e^{i\omega t}$, leading to the eigenvalue equation~\cite{Nori2019, Kuzyk2018}
\begin{equation} \label{EigenvalueModes}
    L[\mathbf{U}] = -\omega^2 \mathbf{U},
\end{equation}
where $L[\mathbf{U}]  = \rho^{-1}(\lambda_0 + \mu)\nabla (\nabla \cdot \mathbf{U(\mathbf{r}})) + \mu \nabla^2 \mathbf{U}(\mathbf{r})$. The above eigenvalues equation is challenging due to the geometry of the device and initial conditions. Here, $\mathbf{U}$ is the displacement vector, which must be a linear combination of a series of eigenstates for the equation, called modes \cite{Ph3}.

The eigenvalue equation~\eqref{EigenvalueModes} is a partial differential equation (PDE) of second order for a system with several possible modes that can be numerically solved using the finite elements approximation (FEA) method~\cite{Ph2}. FEA is applied to solve this kind of PDE by generating a mesh that discretizes the structure and calculating an approximated and discrete solution for this mesh. A FEA simulation was implemented using the Partial Differential Equation Toolbox of MATLAB to calculate the phonon profile of the diamond structure. This toolbox can get approximated results for PDE systems applied in problems like heat transfer or structural analysis~\cite{Ph1}. \par 

The following pseudo-code illustrates how to find the vibrational modes:

\begin{algorithmic}[1]
\State Define geometry and mesh grid
\State Create a circular shape and replicate it across the domain
\State Remove duplicate points and construct a structural model
\State Plot and save geometry visualization
\State Set structural properties (Young's modulus, Poisson's ratio, density)
\State Apply boundary conditions on specified faces
\State Generate and save mesh
\State Define frequency range and solve for natural frequencies
\State Plot and save frequency results
\State Identify and plot modes near 50 GHz
\end{algorithmic}

\bibliographystyle{unsrt}

\end{document}